\documentclass[aps,prx,twocolumn,groupedaddress]{revtex4-2}
\usepackage{graphicx}
\usepackage{amsmath}
\usepackage{amssymb}
\usepackage{siunitx}
\usepackage{color}
\usepackage{ulem}

\newcommand{\e}{\mathrm{e}}

\newcommand{\vac}{\varnothing}
\newcommand{\ket}[1]{|#1\rangle}
\newcommand{\bra}[1]{\langle#1|}

\begin{document}

%\preprint{}

\title{Collective strong coupling in a plasmonic nanocavity}
\author{H. Varguet}
\author{A. A. D\'iaz-Valles}
\author{S. Gu\'erin}
\author{H.R. Jauslin}
\author{G. {Colas des Francs}}
\affiliation{Laboratoire Interdisciplinaire Carnot de Bourgogne (ICB), UMR 6303 CNRS, Universit\'e Bourgogne Franche-Comt\'e,
9 Avenue Savary, BP 47870, 21078 Dijon Cedex, France}
\email{gerard.colas-des-francs@u-bourgogne.fr}

\begin{abstract}
Quantum plasmonics extends cavity quantum electrodynamics (cQED) concepts to the nanoscale, taking benefit from the strongly subwavelength confinement of the plasmon modes supported by metal nanostructures. In this work, we describe in detail collective strong coupling to a plasmonic nanocavity. Similarities and differences to cQED are emphasized. We notably observe that the Rabi splitting can strongly deviate from the standard $\sqrt{N_e}\Delta \Omega_1$ law, where $N_e$ is the number of emitters and $\Delta \Omega_1$ the Rabi splitting for a single emitter. In addition, we discuss the collective Lamb shift and the role of quantum corrections to the emission spectra. 
\end{abstract}
\date{\today}

%\keywords{}

\maketitle

\section{Introduction}
Light emission by dipolar quantum emitters (QEs) strongly depends on their environment. It can be controlled by placing emitters in a designed cavity, which is the basis of cavity quantum electrodynamics (cQED) concepts \cite{Vahala:2003}. For an emitter weakly coupled to the cavity, the decay rate of the excited state can be enhanced to reach the coherence limit and produce indistinguishable photons of strong interest for quantum technologies \cite{Somaschi-Senellart:16}. In the strong coupling regime, there is reversible exchange of energy  between emitters and cavity so that hybridization of their states occurs, leading to the so-called dressed states because the emitter's states are modified in presence of the cavity. The manipulation of dressed states permits the generation of non classical photon states \cite{Faraon-Vukovic:08} but also to control chemical reactions \cite{Shalabney-Ebbesen:2015,Kato:18,Herrera-Owrutsky:20,Flick-Narang:20}, opening the door towards exciting and unexpected applications \cite{Flick:18}. Moreover, strong efforts have been made since more than a decade to transpose cQED concepts to nanophotonics and plasmonics for integration purposes \cite{Hummer-Garciavidal:2013,Torma-Barnes:2015,Ginzburg:16,Vasa-Lienau:17,Bozhevolnyi-Khurgin:17,LalanneReview:18,Cortes-Gray:20}. The strong coupling regime is characterized by the Rabi splitting in the spectrum of the coupled system that generally follows $\hbar \Omega_R \propto \hbar \sqrt{N_e/V_m}$ where $N_e$ is the number of QEs and $V_m$ is the effective volume of the cavity mode involved in the coupling process \cite{Zengi-Kall-Shegai:2015,Nabiev:18}. Quantum plasmonics mainly relies on the strongly subwavelength plasmon volume to achieve the strong coupling regime but collective strong coupling can further increase the effect. 

In this work, we analyse in detail the collective strong coupling regime in a plasmonic nanocavity. We discuss the role of the localized surface plasmons (LSPs) in the coupling process and discuss the dependence of the Rabi splitting on the number of QEs. We also carefully investigate quantum corrections to the collective Lamb shift and observe a significant effect on the dressed states energies.

\section{Quantum description of the coupled system}
\begin{figure}[h]
	\includegraphics[width=4cm]{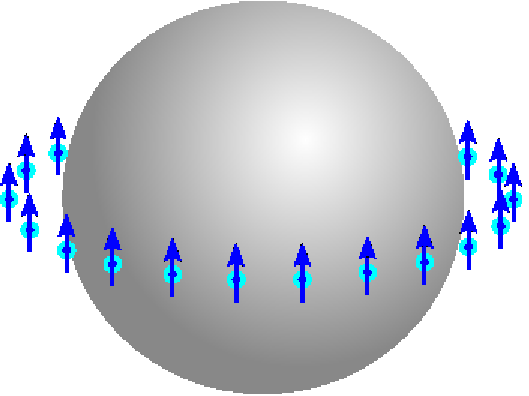}
	\caption{Spatial configuration for the strong coupling regime. $N_e$ orthoradial QEs are homogeneously spread in the equator plane 2 nm from the silver particle of radius $R=8$ nm. The dipole moment is $d=24$ D.
\label{Fig:config}}
\end{figure}

In this section, we briefly recall the main ingredients of the model and state our main working hypothesis. A scheme of the hybrid system is shown in Fig. \ref{Fig:config}. $N_e$ QEs are located close to a spherical metal nanoparticle (MNP) so that they can collectively couple to the LSPs supported by the MNP.  
\subsection{Lossy plasmon modes quantization and continuous model}
The Hamiltonian of $N_e$ identical emitters coupled to the MNP is
\begin{eqnarray}
\label{hamil}
\hat H&=&\hat H_{\mathrm{QE}}+\hat H_{\mathrm{MNP}}+\hat H_\mathrm{I} \,,\\
\nonumber 
\hat H_{\mathrm{QE}}&=&\sum_{j=1}^{N_e}\hbar\left(\omega_0-i\frac{\gamma_0}{2}\right)\hat{\sigma}_{+}^{(j)}\hat{\sigma}_{-}^{(j)} \,,\\
\nonumber 
\hat H_{\mathrm{MNP}}&=&\int d\mathbf{r} \int_0^{+\infty}\!\!\!\!\!\!\! d\omega\ \hbar\omega \hat{\mathbf{f}}_\omega^{\dagger}(\mathbf{r})\cdot\hat{\mathbf{f}}_\omega(\mathbf{r}) \,,\\
\nonumber 
\hat H_\mathrm{I}&=&-\sum_{j=1}^{N_e}\left[\hat{\sigma}_{+}^{(j)} \otimes \int_0^{+\infty}\!\!\!\!\!\!\! d\omega\ \mathbf{d}^{(j)}\cdot\hat{\mathbf{E}}^{+}_\omega(\mathbf{r}_j)+H.c.\right] \;.
\end{eqnarray}
$\omega_0$ is the transition angular frequency between the ground state $\ket{g}_j$ and excited state $\ket{e}_j$ of emitter $j$ and we introduce the coupling operator of the $j^{th}$ emitter $\hat{\sigma}_{+}^{(j)}=\ket{e}_j ~_j\bra{g}$ and  $\hat{\sigma}_{-}^{(j)}=\ket{g}_j~_j\bra{e}$. In equation (\ref{hamil}), the first term is the QE energy and we have phenomelogically introduced the decay rate $\gamma_0$ of the excited state. The second term describes  the total energy of the electromagnetic field where $\hat{\mathbf{f}}^{\dagger}({\mathbf r})$ ($\hat{\mathbf{f}}({\mathbf r})$) is the LSP polaritonic vector field operator at the position ${\mathbf r}$ associated to the creation (annihilation) of a quantum in the presence of the MNP. The last term describes the emitters-field interaction under the rotating-wave approximation.We discuss later this approximation. 

$\hat{\mathbf{E}}^{+}_\omega$ is the electric field operator associated to the field scattered in presence of the MNP. We note $\mathbf{G}_{tot}({\mathbf r},{\mathbf r}')$ the Green tensor associated to the electric field response at position ${\mathbf r}$ from an excitation localized at ${\mathbf r}'$ in the medium. The electromagnetic field is quantized within the Langevin type model \cite{Knoll-Scheel-Welsch:01} and the electric field operator can be written as
\begin{eqnarray}
\mathbf{\hat{E}}^{+}_\omega(\mathbf{r})=& i\sqrt{\frac{\hbar}{\pi\epsilon_0}}k_0^2
\int d{\mathbf{r}'} \sqrt{\varepsilon''_\omega(\mathbf{r}')}
{\mathbf G}_{tot}({\mathbf r},{\mathbf r}',\omega)\hat{{\mathbf f}}_\omega({\mathbf r'}) \hspace{0.5cm}
\label{eq:OpE}
\end{eqnarray}
where $k_0=\omega/c$ is the wavenumber.  
The wave function of the hybrid system can be written at time $t$ as
\begin{eqnarray}
\label{wavefun}
&&\ket{\psi(t)} =\sum_{j=1}^{N_e} C_{e,\vac}^{(j)} (t)\ket{e^{(j)},\varnothing}\\
\nonumber
&&
+ \int d\mathbf{r}\int_0^{+\infty}d \omega\ e^{-i\omega t}\mathbf{C}_{g}(\mathbf{r},\omega,t)\cdot\ket{g,\mathbf{1}_\omega(\mathbf{r})}  \;,
\end{eqnarray}
where $\ket{e^{(j)},\vac}$ represents the $j^{th}$ emitter in its excited state and no LSP mode excited. For instance, for $N_e=4$ emitters, $\ket{e^{(2)},\vac}=\ket{g,e,g,g,\vac}$. $\ket{g,\mathbf{1}_\omega(\mathbf{r})}$ corresponds to all emitters in their ground state and a single excited LSP mode of energy $\hbar \omega$. For the collective strong coupling, we consider the initial coherent superposition 
\begin{eqnarray}
\ket{\psi(0)}=\ket{B,\vac}=\frac{1}{\sqrt{N_e}}\sum_{j=1}^{N_e}\ket{e^{(j)},\vac} \,.
\end{eqnarray}
This state corresponds to the last excited state in the Dicke superradiance cascade \cite{Varguet-GCF:19} but an initial state obtained by incoherent pumping can be considered as well \cite{Delga-GarciaVidal:2014}.
The optical response is characterized by the spectral density 
\begin{eqnarray}
\label{eq:spec}
D(\omega)&=&\frac{\gamma_0}{2\pi}\left \vert \int_0^{+\infty} C_{B,\vac}(t)e^{i\omega t} dt \right\vert^2 \\
&=&\frac{\gamma_0}{2\pi N_e}\left \vert \sum_{j=1}^{N_e} \int_0^{+\infty} C_{e,\vac}^{(j)} (t)e^{i\omega t} dt \right\vert^2 
\nonumber
\end{eqnarray}
where $C_{B,\vac}(t)=\bra{B,\vac} \psi(t)\rangle$ is the probability amplitude associated to the state $\ket{B,\vac}$.
The dynamics of the wavefunction follows the time-dependent Schr{\"o}dinger equation 
\begin{eqnarray}
\label{eq:schrodinger}
i\hbar \frac{\partial}{\partial t} \ket{\psi(t)}=\hat H \ket{\psi(t)}
\end{eqnarray}
so that the evolution of the spectral density  obeys (see appendix \ref{sectSupp:ContSpec})
\begin{eqnarray}
\nonumber
&&D(\omega) \propto \left \vert \sum_{k=1}^{N_e}\sum_{l=1}^{N_e}\left\{ \left[ \left( (\omega-\omega_0)+i\frac{\gamma_0}{2}\right)\mathbb{\mathbf{I}}+\mathbf{M}(\omega)\right]^{-1}\right\}_{kl} \right \vert^2 \,,\\
\label{eq:SpecCont}
&&M_{ij}(\omega)=i\frac{\omega^2}{\hbar \epsilon_0c^2}Im[G_{ij}(\omega)] \\
\nonumber
&& \hspace{2cm} +\frac{1}{\hbar \pi \epsilon_0} {\mathbb P} \int_0^{\infty} d\omega' \frac{\omega'^2}{c^2} \frac{Im[G_{ij}(\omega')]}{\omega'-\omega}\\
&&G_{ij}(\omega)=\mathbf{d}^{(i)} \cdot \mathbf{G}_{tot}(\mathbf{r}_i,\mathbf{r}_j,\omega)\cdot \mathbf{d}^{(j)} \;.
\nonumber
\end{eqnarray}

We decompose the Green tensor $\mathbf{G}_{tot}=\mathbf{G}_0+\mathbf{G}_{scatt}$ where $\mathbf{G}_0$ is the free space contribution (homogeneous background of dielectric constant $\epsilon_b$) and $\mathbf{G}_{scatt}$ refers to the MNP contribution only. We emphasize that we phenomenologically introduced the free-space decay rate $\gamma_0$ in the Hamiltonian (eq. \ref{hamil}),  assuming the free space  Lamb shift  taken into account into $\omega_0$. Since the free-space contribution is taken into account in $\gamma_0$ and $\omega_0$, $G_{ii}(\omega)=\mathbf{d}^{(i)} \cdot \mathbf{G}_{scatt}(\mathbf{r}_i,\mathbf{r}_i,\omega)\cdot \mathbf{d}^{(i)}$ for $j=i$. See also Refs. \cite{Dung-Knoll-Welsch:2002,DrezetPRA:16,DrezetPRA:17,Dorier-Jauslin:19,VarguetFano:19,Dorier:20} for a discussion on the free-space contribution.

We will discuss in detail the interpretation of the term $M_{ij}$ as the dipole-dipole shift \cite{Fleischhauer:2011,Tian-Zhao:19} in section \ref{sect:Lamb}. We will notably demonstrate the role of the quantum correction, corresponding to the integral term in the expression of $M_{ij}$. This term is generally neglected for single emitter configurations \cite{vanVlack-Hughes:2012,Delga-Feist-GarciaVidal:14,Varguet-GCF:16} but we will show that it is proportionnal to the number of emitters for all emitters at the same location. Practically, the numerical simulations of the spectral density necessitate to compute the quantum correction integral in $M_{ij}$ that could lead to numerical difficulties due to the pole on the real axis. Better convergence can be achieved deforming the integration path on the imaginary axis \cite{Fleischhauer:2011} or using a symmetry relation of the Green function to remove the singularity \cite{Tian-Zhao:19}. Still, it necessitates numerical integration. An alternative approach consists in the effective model described in the next section.

\subsection{Effective model}
\label{sect:Heff}
The construction of the effective model relies on the Lorentzian profile of the LSP resonances \cite{Dzsotjan-GCF:2016}. We use a modal Mie expansion of the Green tensor $\mathbf{G}_{scatt}=\sum \mathbf{G}_{scatt}^n$ and the coupling constant between $j^{th}$ QE and LSP$_n$ obeys    
\begin{eqnarray} 
\nonumber
\vert \kappa_{n}^{(j)}(\omega)\vert^2 &=&\frac{k_0^2}{\hbar\pi\epsilon_0} Im\left[\mathbf{d}^{(j)} \cdot \mathbf{G}_{scatt}^n(\mathbf{r}_j,\mathbf{r}_j,\omega)\cdot \mathbf{d}^{(j)}\right]\\ 
&=&\frac{\Gamma_n}{2\pi}\frac{[g_n^{(j)}]^2}{(\omega-\omega_n)^2+\frac{\Gamma_n^2}{4}} \;.
\label{eq:Kappa}
\end{eqnarray}
with the coupling strength $g_n^{(j)}$ to the plasmon mode of order n (LSP$_n$). $\omega_n$ and $\gamma_n$ are the LSP$_n$ resonances frequencies and widths, respectively. An effective Hamiltonian is derived \cite{Dzsotjan-GCF:2016}
\begin{eqnarray}
\label{eq:Heff1}
&&\hat H=\hat H'_{\mathrm{QE}}+\hat H_{\mathrm{LSP}}+\hat H_{\mathrm{QE-LSP}} \,,\\
\nonumber 
&&\hat H'_{\mathrm{QE}}=\sum_{j=1}^{N_e}\left(\omega_0+\delta\omega_0^{(j)}-i\frac{\gamma_0}{2}\right)\hat{\sigma}_{+}^{(j)}\hat{\sigma}_{-}^{(j)} \,,\\
\nonumber
&&\delta\omega_0^{(j)}=-\frac{k_0^2}{\hbar\epsilon_0}\sum_{i=1}^{N_e}Re\left[\mathbf{d}^{(j)} \cdot \mathbf{G}_{0}(\mathbf{r}_j,\mathbf{r}_i,\omega)\cdot \mathbf{d}^{(i)} \right]
\\
\nonumber 
&&\hat H_{\mathrm{LSP}}=\sum_{j=1}^{N_e}\sum_{n=1}^{N}\hbar(\omega_n-i\gamma_n)\hat{a}_n^{(j)\dagger}\hat{a}^{(j)}_n \,,\\
\nonumber 
&&\hat H_{\mathrm{QE-LSP}}=\hbar\sum_{j=1}^{N_e} \sum_{n=1}^{N}\left(g^{(j)}_n\hat{\sigma}^{(j)}_+\hat{a}^{(j)}_n+(g_n^{(j)})^*\hat{a}_n^{(j)\dagger}\hat{\sigma}^{(j)}_-\right) \;.
\end{eqnarray}
$\delta\omega_0^{(j)}$ is frequency shift induced by the free-space dipole-dipole Van der Waals interactions between the atoms \cite{Gross-Haroche:1982,Delga-GarciaVidal:2014}. We don't include the free-space dipole-dipole coupling (imaginary part) since it is negligible.
$\hat{a}_n^{(j)\dagger}$ and $\hat{a}_n^{(j)}$ describe the creation or annihilation of a plasmon of order $n$ by the emitter $j$. Building the effective Hamiltonian, we have defined LSPs with respect to the position of the emitter that can excite them, leading to non-orthogonal modes. As a consequence, $\hat{a}^{(j)}_n$ operators do not obey the standard bosonic commutation relations but 
\begin{eqnarray}
&&\left[\hat{a}_n^{(i)},\hat{a}_{n' }^{(j)\dagger}\right]=\delta_{nn'}\mu_n^{(ij)},\label{com} \\
&&\mu_n^{(ij)}=\frac{1}{\pi \hbar\varepsilon_0}\frac{\omega_n^2}{c^2}\frac{ \mathfrak{Im}\left[\mathbf{d}^{(i)}\cdot\mathbf{G_n}(\mathbf{r}_i,\mathbf{r}_j,\omega_n)\cdot \mathbf{d}^{(j)}\right]}{\kappa_n^{(i)}(\omega_n)\kappa_n^{(j)^*}(\omega_n)} \;,
\label{eq:muij}
\end{eqnarray}
where $\mu_n^{(ij)}$ is the modal overlap function. 
One can define $N_{ind}$ independent creation ($\hat{b}_n^{(l)^\dagger}$) and annihilation ($\hat{b}_n^{(l)}$) operators satisfying the standard commutation rules by a L\"owdin orthormalisation \cite{Annavaparu:13,Castellini:18,Varguet-GCF:19}. This leads to an alternative form of the effective Hamiltonian 
\begin{eqnarray}
\label{eq:Heff2}
&&\hat H=\hat H'_{\mathrm{QE}}+\hat H'_{\mathrm{LSP}}+\hat H'_{\mathrm{QE-LSP}} \,,\\
\nonumber 
&&\hat H'_{\mathrm{LSP}}=\sum_{n=1}^{N}\sum_{j=1}^{N_{ind}}\hbar\left(\omega_n-i\frac{\gamma_n}{2}\right)\hat{b}_n^{(j)^\dagger}\hat{b}_n^{(j)} \,,\\
\nonumber 
&&\hat H'_{\mathrm{QE-LSP}}=\sum_{j=1}^{N_e}\sum_{n=1}^{N}\sum_{l=1}^{N_{ind}}\hbar\left(g_n^{(jl)}\hat{\sigma}_{+}^{(j)}\hat{b}_n^{(l)}+H.c\right) \;.
\end{eqnarray}
with the coupling strength $g_n^{(jl)}$ between the emitter $j$ and the mode LSP$_n^{(l)}$. 

The wave function of the hybrid system is finally  written on a single excitation basis of dimension $M=N_e+N\times N_{ind}$ as
\begin{eqnarray}
\label{wavefunEff}
\ket{\psi(t)} =\sum_{j=1}^{N_e} C_{e,\vac}^{(j)} (t)\ket{e^{(j)},\varnothing}&& \\
\nonumber
&&+\sum_{n=1}^{N} \sum_{l=1}^{N_{ind}}C_{g,1}^{n,(l)} (t)\ket{g,1_n^{(l)}}\;.
\end{eqnarray}
$\ket{1_n^{(l)}}=\hat{b}_n^{(l)^\dagger}\ket{\varnothing}$ refers to a L{\"o}wdin plasmonic state LSP$_n^{(l)}$.

The dynamics of the hybrid system is efficiently described in the eigenbasis of the effective Hamiltonian; the $M$ eigenvectors (dressed states) $\ket{\Pi_m^D}$ are associated to the complex eigenvalues $\Lambda_m=\Omega_m-i\Gamma_m/2$ and 
\begin{eqnarray}
&&\ket{\psi(t)} =\sum_{m=1}^{M} \eta_m e^{-i\Lambda_m t} \ket{\Pi_m^D} \,,\\
&&\eta_m=\bra{\Pi_m^D}\psi(0)\rangle \;.
\nonumber
\end{eqnarray}

Assuming again the initial state $\ket{\psi(0)}=\ket{B,\vac}$ defined above, the spectral density is finally expressed as 
\begin{eqnarray}
\label{eq:SpecHeff}
&&D(\omega)=\frac{\gamma_0}{2\pi N_e^2} \left \vert \sum_{m=1}^M\frac{\vert\sum_{l=1}^{N_e} m_0^{(l)}\vert^2 }{\omega-\Omega_m+i\frac{\Gamma_m}{2}} \right\vert^2 \\
&&m_0^{(l)}=\bra{e^{(l)},\vac}\Pi_m^D\rangle
\nonumber
\end{eqnarray}

\section{Collective strong coupling}

\subsection{Rabi splitting as a function of the number of emitters}
\begin{figure*}\includegraphics[width=0.97\textwidth]{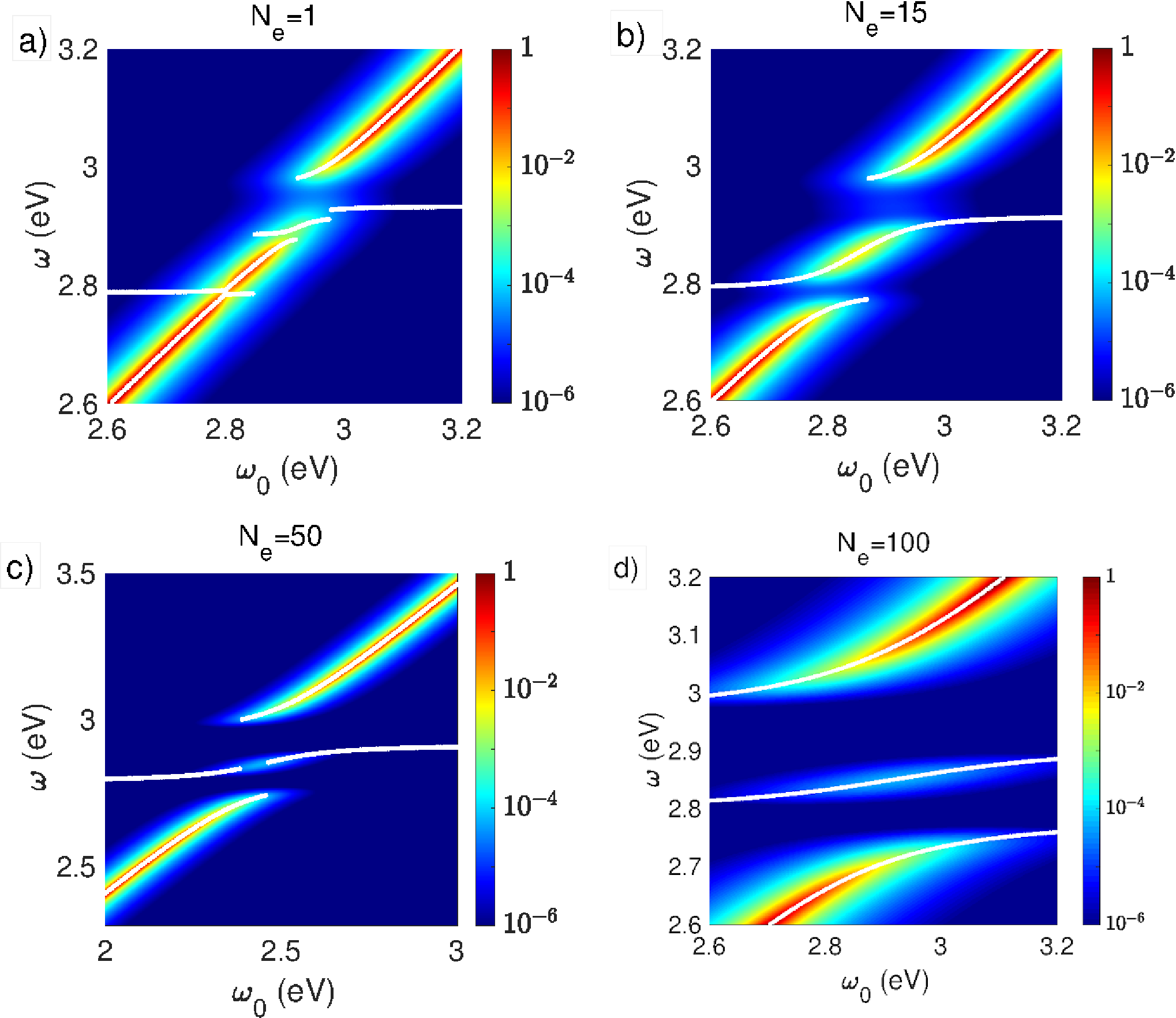}
	\caption{Normalized emission spectra for (a) $N_e=1$, (b) $N_e=15$, (c) $N_e=50$ and (d) $N_e=100$ (the Van der Waals frequency shift $\delta \omega_0$ is ignored for $N_e=100$, see the text for detail). The color maps corresponds to the density spectra calculated using Eq. (\ref{eq:SpecHeff}) and solid white lines indicate the angular frequencies $\Omega_m$ of the dressed states that contribute mainly to the strong coupling. The emission spectra are normalized with respect to their maximum. 
\label{Fig:NearFieldSpecHeff}}
\end{figure*}

%\subsubsection{Emission spectrum}
Emission spectra as a function of the QE transition frequency $\omega_0$ are presented in Fig. \ref{Fig:NearFieldSpecHeff} for $N_e=$1 to 100 QEs. Van der Waals dipole-dipole interactions induce a blue shift that strongly depends on the distance $d_{12}$ between two adjacent dipoles and ranges from $\delta \omega_0=12$ meV for 15 atoms ($d_{12}=4.2$ nm),  $\delta \omega_0=438$ meV (50 atoms, $d_{12}=1.3$ nm)  to $\delta \omega_0=3.5$ eV ($d_{12}=0.6$ nm). However, the dipole-dipole interaction is probably overestimated for $d_{12}=0.6$ nm and a dedicated model should be considered, beyond the scope of this work \cite{Eberlein:04}. Since this solely induces a frequency shift that is identical for all the emitters, but does not change the Rabi splitting value, we ignore it for $N_e=100$, without loss of understanding the strong coupling characteristics.

For a single emitter, we observe a Rabi splitting $\Delta \Omega_1=88$ meV, for a QE emission close to the high order LSP resonances  $\omega_0\approx \omega_\infty\approx 2.95$ eV, originating from the coupling to numerous high order modes \cite{Varguet-GCF:16}. The Rabi splitting increases as a function of the number of QEs and reaches $\Delta \Omega_{100}=217$ meV for $N_e=100$. For $N_e>15$, we also observe the opening of a second gap at QE emission close to the dipolar LSP$_1$ resonance ($\omega_0\approx \omega_1 \approx 2.8$ eV). This results from the collective coupling of all $N_e$ emitters to the dipolar LSP$_1$ mode. The LSP$_1$ mode radiatively leaks into the far-field so that such strong coupling regime can be recorded in the far-field zone \cite{Zengi-Kall-Shegai:2015,Rousseaux_Johansson:20}. For $N_e\sim 50$ the two strong coupling regimes occur at the same QE emission frequency $\omega_0 \approx 2.85$ eV and the two Rabi splittings become difficult to separate for higher number of QEs. See also the discussion in section \ref{sect:Dressed}.
\\
	
It is of great interest to compare the collective strong coupling behaviour to the ideal situation where all the emitters are located at the same position. In this case all the $N_e$ QEs are coupled to the same L{\"o}wdin plasmon modes LSP$_n^{(1)}$. All the emitters are fully equivalent and one can define collective operators \cite{Varguet-GCF:19} 
\begin{eqnarray}
&& \hat S_+=\frac{1}{\sqrt{N_e}}\sum_{j=1}^{N_e}\hat \sigma_+^{(j)} \;, \\
&& \hat S_+=\hat S_- ^\dagger \,,
\end{eqnarray}
in order to display the collective coupling $\sqrt{N_e} g_n$ in the interaction Hamiltonians. Indeed, $g_n^{j1} =g_n \forall j$ and   $g_n^{jl} =0$ if $l\ne 1$, so that the effective Hamiltonian associated to the ideal configuration simplifies to the following representation in the basis $\left\{ \ket{B,\vac}, \ket{g,1_1^{(1)}},\ket{g,1_2^{(1)}},\ldots,\ket{g,1_N^{(1)}}\right\}$

\begin{figure}
\includegraphics[width=0.47\textwidth]{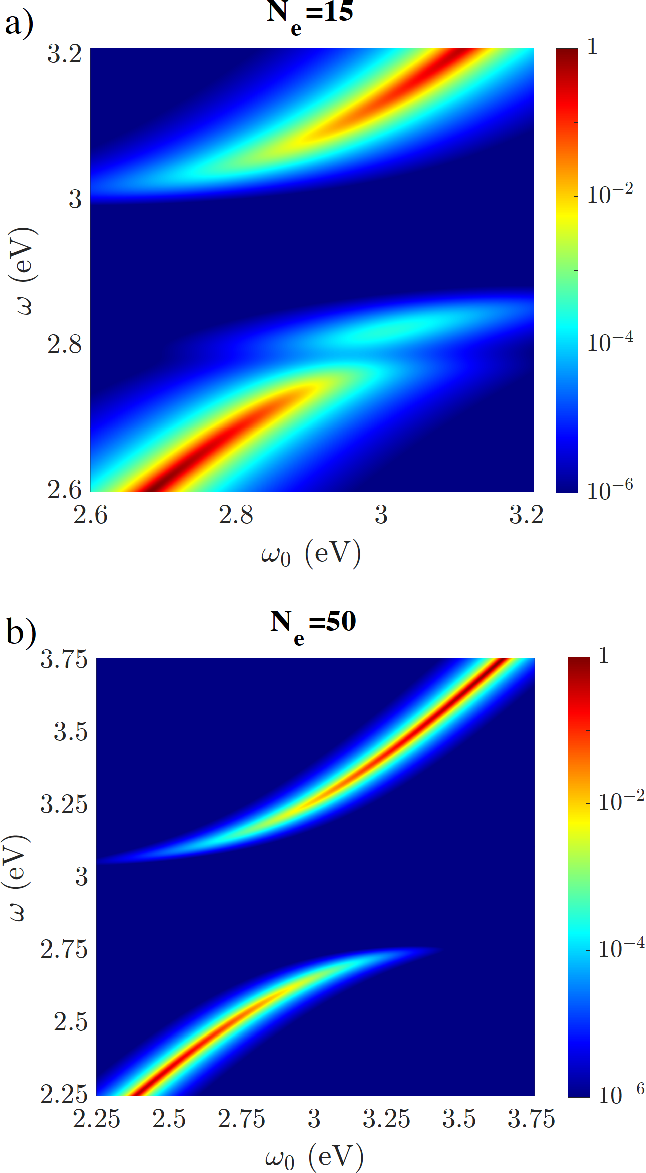}
	\caption{Normalized emission spectra for (a) $N_e=15$, (b) $N_e=50$ considering ideal configuration.
\label{Fig:NearFieldSpecHeffIdeal}}
\end{figure}
\begin{figure}
	\includegraphics[width=0.47\textwidth]{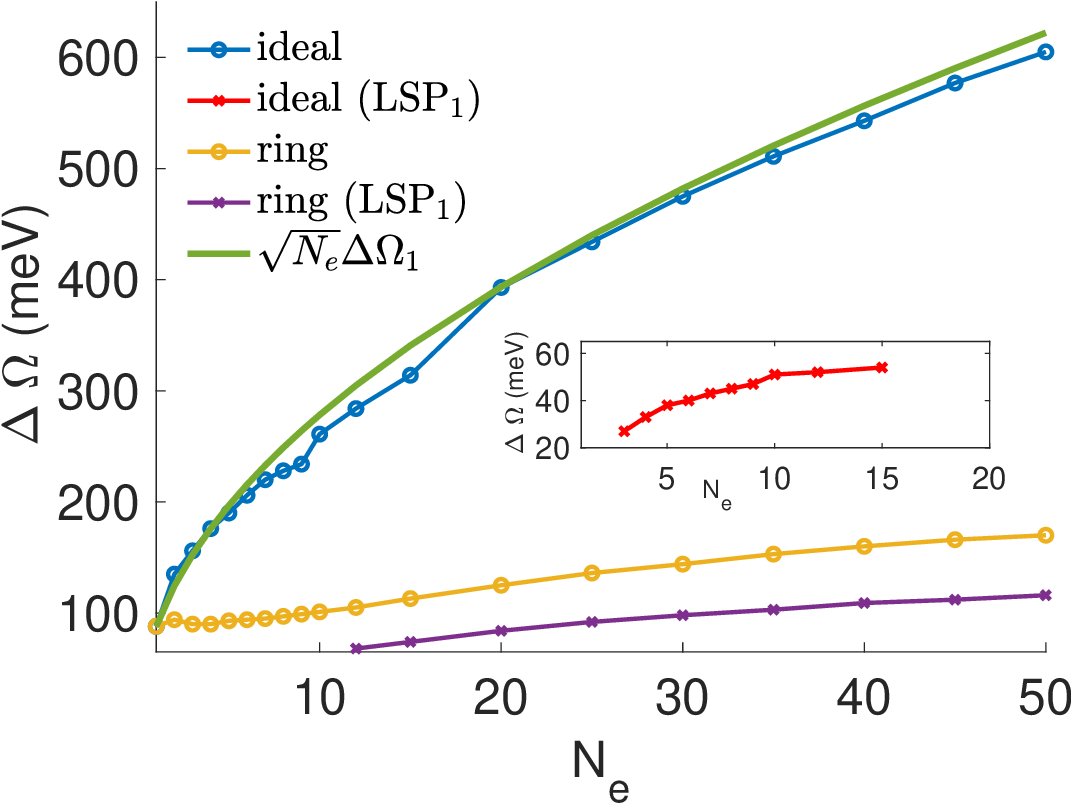}
	\caption{Rabi splitting as a function of the number of emitters. The blue curve corresponds to all QEs at the same position (ideal configuration) and follows a $\sqrt{N_e} \Delta\Omega_1$ law (green curve). The yellow and purple curves represents the Rabi splitting associated to the strong coupling to high-oder mode or dipolar mode, respectively. The inset represents the evolution of the Rabi splitting between emitters and LSP$_1$ in the ideal configuration.  
\label{Fig:RabiSplit}}
\end{figure}

	
\begin{eqnarray}
\label{eq:HeffIdeal}
H_{\mathrm{eff}}=\hbar\begin{bmatrix}\omega_0-i\frac{\gamma_{0}}{2} & \sqrt{N_e}g_1 &  \sqrt{N_e}g_2 & \cdots &  \sqrt{N_e}g_N\\
 \sqrt{N_e}g_1 & \omega_1-i\frac{\gamma_1}{2} & 0 & \cdots & 0\\
 \sqrt{N_e}g_2 & 0 & \omega_2-i\frac{\gamma_2}{2} & \ddots & \vdots\\
\vdots & \vdots & \ddots & \ddots & 0\\
 \sqrt{N_e}g_N & 0 & \cdots & 0 & \omega_N-i\frac{\gamma_N}{2}\end{bmatrix} \hspace{0.55cm}
\end{eqnarray}
The emission spectra in the ideal configuration are presented in Fig. \ref{Fig:NearFieldSpecHeffIdeal}. The Rabi splitting is strongly increased compared to the ring configuration (Fig. \ref{Fig:NearFieldSpecHeff}) due to the collective coupling to the same mode. We compare the Rabi splittings obtained from the two configurations as a function of the number of QEs in Fig. \ref{Fig:RabiSplit}. For the ideal configuration, we recover almost the usual $\sqrt{N_e} \Delta\Omega_1$ collective Rabi splitting. The small deviation to the $\sqrt{N_e} \Delta\Omega_1$ law is attributed to the difficulty to separate the strong coupling regime to either the high order LSP mode or the dipolar mode LSP$_1$.  See the next section for a discussion. The Rabi splitting becomes much lower for the ring configuration of QEs (see also ref. \cite{Antosiewicz:19}). This originates from a discernable coupling of each emitter to LSP modes, strongly jeopardizing the collective behaviour. A similar blockade of the collective coupling has been discussed in the weak coupling regime for plasmonic superradiance \cite{Varguet-GCF:19}.

At this point, we can discuss the rotating wave approximation we have done (Eq. \ref{hamil}) and the feasibility of reaching ultra-strong coupling \cite{Kockum-Nori:19}. The atomic transition is $\omega_0\sim 3$ eV and the coupling strength between a single quantum emitter and localized plasmons is typically $g\sim 15$ meV in the discussed configuration. Collective coupling with $N_e$ emitters ideally located at the same position follows $g_{N_e}\sim \sqrt{N_e}g$ so that ultrastrong coupling could be reached for $N_e\sim 400$ emitters ($g_{N_e}\sim 0.1 \omega_0$). For radially oriented emitters, $g\sim 50$ meV \cite{VarguetFano:19} so that the ultrastrong coupling could occur for only $N_e\sim 40$ emitters if located at the same position. For emitters spread all around the particle, the collective coupling is significantly below $\sqrt{N_e}g$, even for the optimized ring configuration, and we estimate that ann ultrastrong regime would necessitate several thousand emitters. That is why we don't consider this regime in this work.  However, this could be achieved experimentally and would be of strong interest to investigate, keeping the counterrotating contributions in the interaction Hamiltonian \cite{Delga-GarciaVidal:2014,Rivera:19}.
\subsection{Dressed states}
\label{sect:Dressed}
\begin{figure*}
	\includegraphics[width=0.97\textwidth]{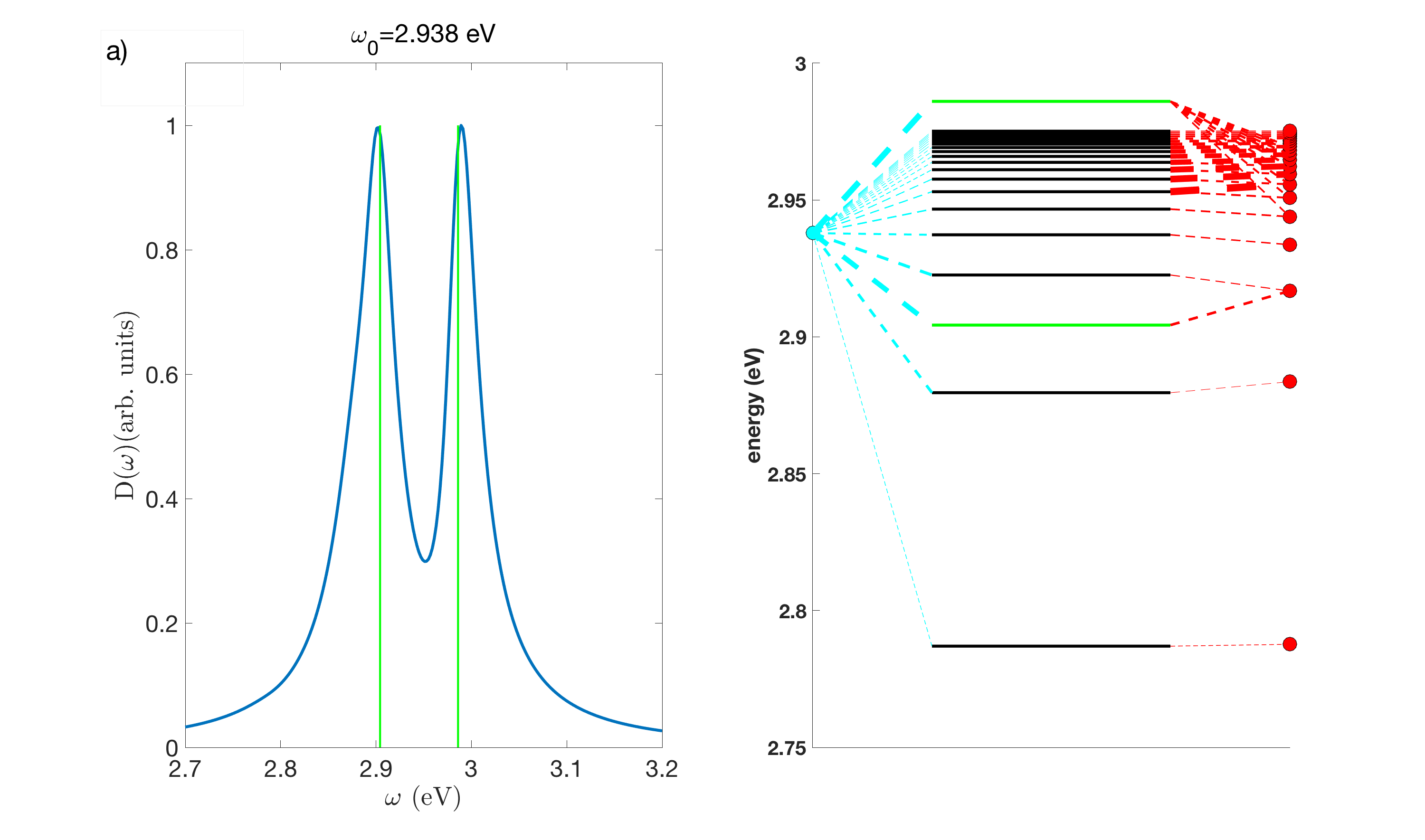}
	\caption{a) Strong coupling spectrum for a single emitter ($N_e=1$). b) Corresponding dressed states. The atomic state energy is represented by the blue point on the left whereas the red points on the right correspond to LSPs energies. Dotted lines represent the states contribution to the hyridized states. The line thicknness is proportionnal to the states' weigth. The two dressed states that mainly contributes to the strong coupling regime are indicated in green. 
\label{Fig:Dressed1}}
\end{figure*}

\begin{figure*}
	\includegraphics[width=0.97\textwidth]{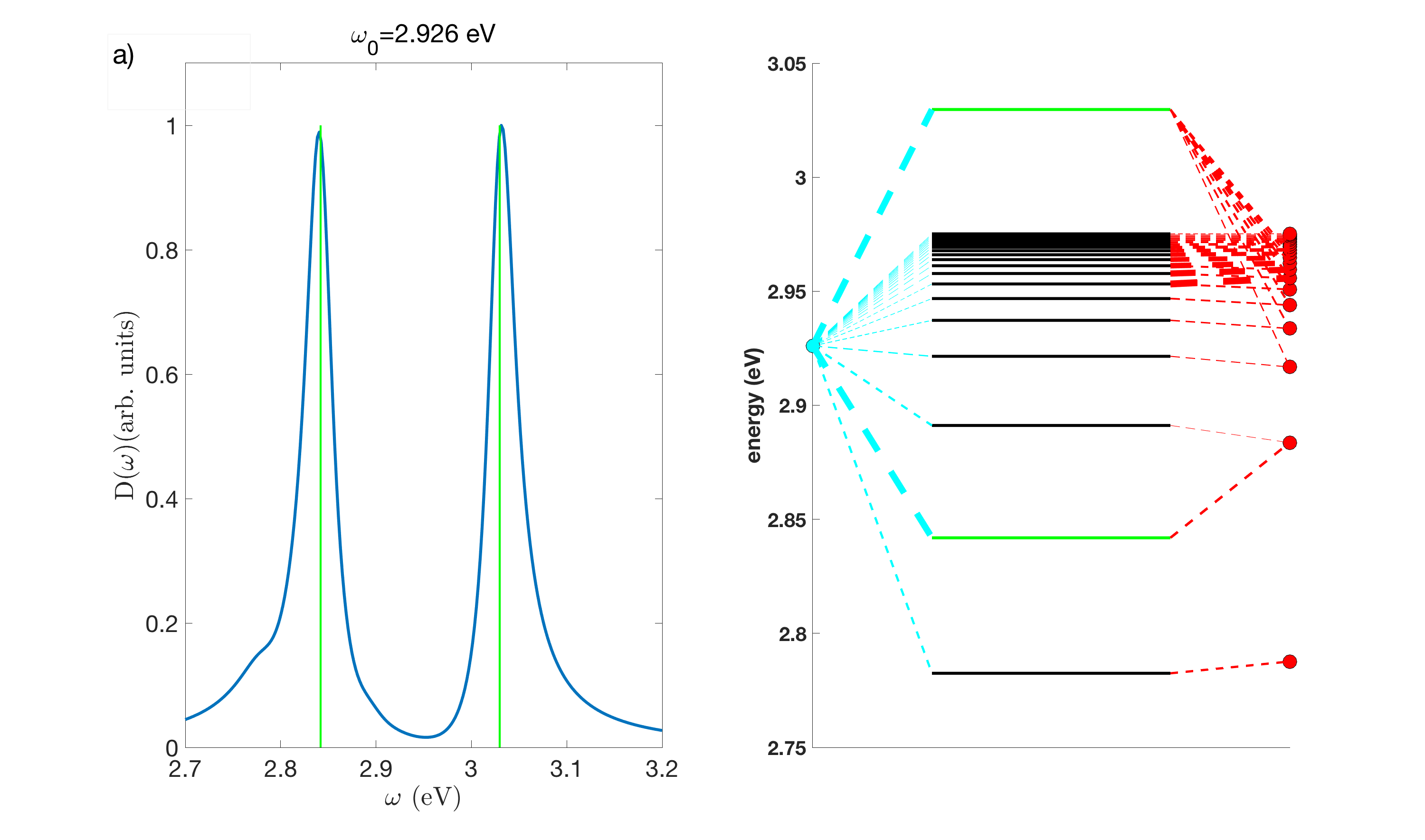}
	\caption{Same as Fig. \ref{Fig:Dressed1} for $N_e=5$.}
\label{Fig:Dressed5}
\end{figure*}

\begin{figure*}
	\includegraphics[width=0.97\textwidth]{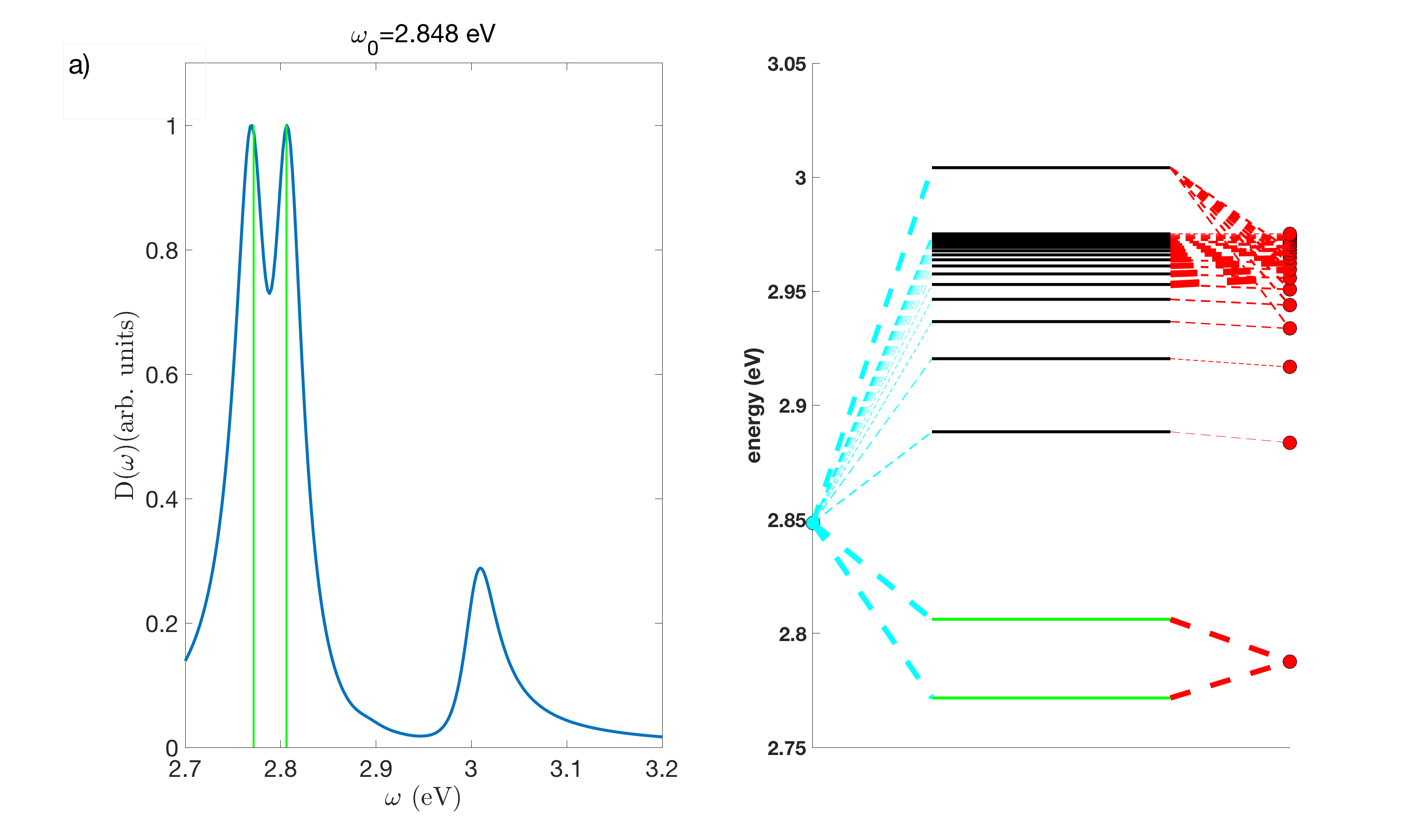}
	\caption{Same as Fig. \ref{Fig:Dressed5} for emitters strongly coupled to LSP$_1$.}
\label{Fig:Dressed5-LSP1}
\end{figure*}

\begin{figure*}
	\includegraphics[width=0.97\textwidth]{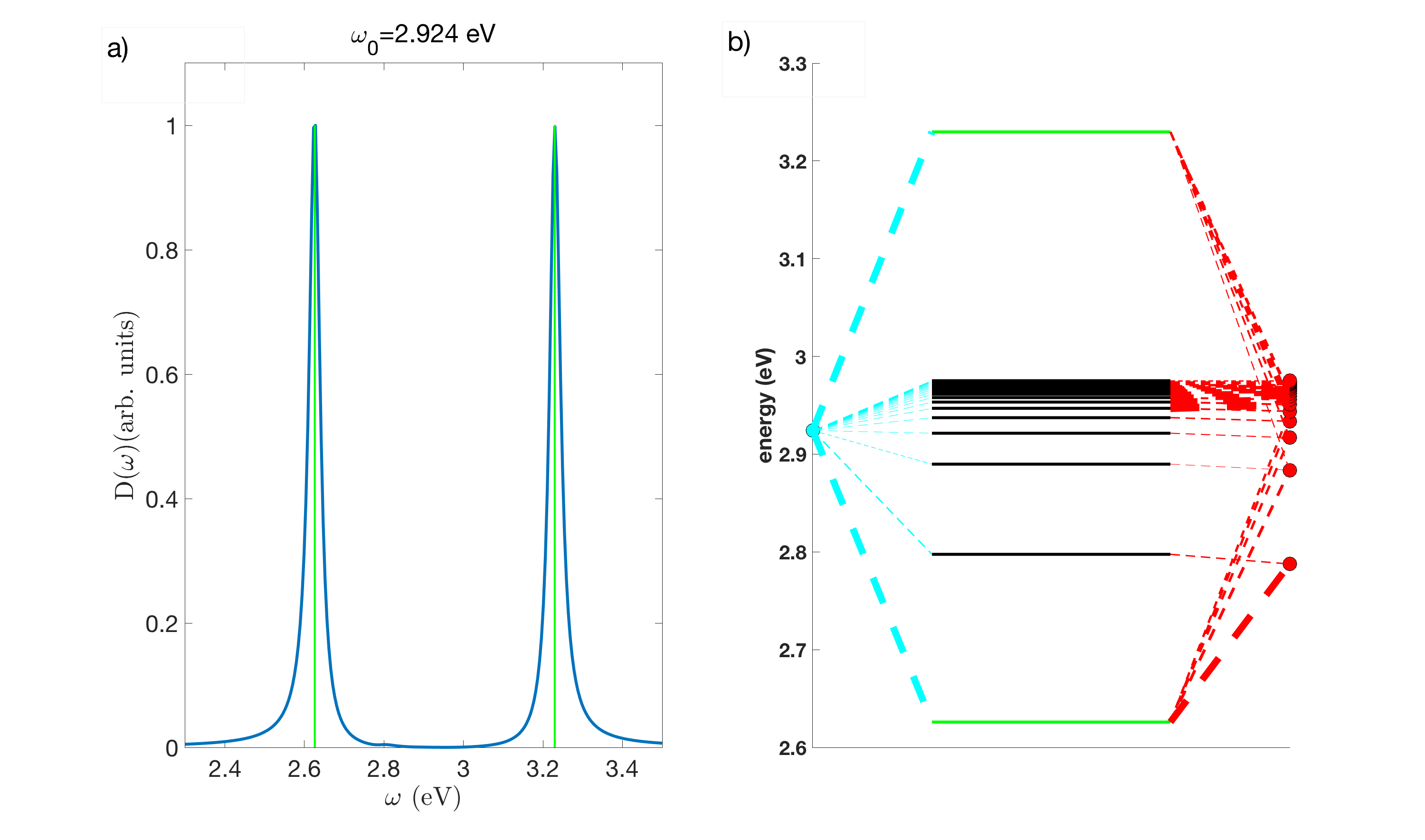}
	\caption{Same as Fig. \ref{Fig:Dressed1} for $N_e=50$.}
\label{Fig:Dressed50}
\end{figure*}

We represent in Fig. \ref{Fig:Dressed1}-\ref{Fig:Dressed50} the evolution of the spectra for $N_e$ emitters strongly coupled to the LSPs modes in the ideal configuration and for $N_e$ ranging from 1 to 50. We also plot the corresponding energy diagrams of the dressed states (Jaynes-Cumming ladder). Dressed states are the eigenvectors of the Hamiltonian so that their energy and the contribution of the emitters and LSPs states are determined from the diagonalization of the Hamiltonian. For $1\le N_e < 3$, the strong coupling regime originates from the cumulative coupling to high order modes (see Fig. \ref{Fig:Dressed1}b and \ref{Fig:Dressed5}b)  \cite{Varguet-GCF:16}. Since high order modes have similar energies, they can be approximated by a single effective LSP mode so that the systems behaves similarly to emitters coupled to a single mode cavity and the Rabi splitting closely follows the $\sqrt{N_e} \Delta\Omega_1$ law .

For $3 \le N_e \le 15$, we also observe a second strong coupling regime with the dipolar LSP mode since only LSP$_1$ is involved (see \ref{Fig:Dressed5-LSP1}b). We estimated in Fig. \ref{Fig:RabiSplit} the Rabi splittings for strong coupling to either high order modes or to LSP$_1$ only.  For $N_e > 15$, the strong coupling occurs with all the LSP modes and a single Rabi splitting can be defined, see Fig. \ref{Fig:Dressed50}b). Finally, we emphasize that the strong coupling regime, including with LSP$_1$, occurs for emitter's frequency $\omega_0$ far from the dipolar mode frequency $\omega_1$, when the number of emitters increases, as clearly observed on Fig. \ref{Fig:Dressed50}b). This originates from the collective Lamb shift discussed in detail in the following section.

\section{Lamb shift and role of quantum corrections}
\label{sect:Lamb}
Additionnal understanding of the collective coupling and the role of the quantum corrections is possible comparing the continuous and effective models. Indeed, the continuous model is fully equivalent to the effective model but it clearly shows quantum corrections to a semi-classical description. In particular, the expression (\ref{eq:SpecCont}) of the spectral density involves the inverse of the matrix $\left[ \left( (\omega-\omega_0)+i\gamma_0/2\right)\mathbb{\mathbf{I}}+\mathbf{M}(\omega)\right]$ with
\begin{eqnarray}
\label{eq:Mij}
&&M_{ij}(\omega)=\Delta \omega_{ij}(\omega)+i\frac{\Gamma_{ij}(\omega)}{2} \\
\nonumber
&&\Gamma_{ij}(\omega)=\frac{2\omega^2}{\hbar \epsilon_0c^2}Im[G_{ij}(\omega)] \\
\nonumber
&& \Delta \omega_{ij}(\omega)=+\frac{1}{\hbar \pi \epsilon_0} {\mathbb P} \int_0^{+\infty} d\omega' \frac{\omega'^2}{c^2} \frac{Im[G_{ij}(\omega')]}{\omega'-\omega}
\end{eqnarray}
so that $\Gamma_{ij}$ and $\Delta \omega_{ij}$ corresponds to the cooperative decay rate modification and dipole-dipole shift  \cite{Knoll-Scheel-Welsch:01,Fleischhauer:2011,VarguetFano:19}. We interprete in detail the dipole-dipole shift expression in the following paragraphs. 

\subsection{Lamb shift}
\subsubsection{Effective model}
For $i=j$, the dipole-dipole shift reduces to the Lamb shift of emitter $j$ \cite{Knoll-Scheel-Welsch:01,Fleischhauer:2011,Tian-Zhao:19}
\begin{eqnarray}
\Delta\omega_{jj}  (\omega_0)&=&\frac{1}{\pi\hbar \epsilon_0} {\mathbb P}\int_{0}^{+\infty} d\omega\frac{\omega ^2}{c^2} \frac{Im G_{jj}(\omega)}{\omega-\omega_0}
\end{eqnarray} 
Recalling the construction of the effective model relies on the Lorentzian profile of the LSP resonances, see Eq. (\ref{eq:Kappa}), the integral can be calculated explicitely and leads to 
\begin{eqnarray}
\Delta\omega_{jj}  (\omega_0)&=&\sum_{n=1}^{N}\frac{g_n^2(\omega_0-\omega_n)}{(\omega_0-\omega_n)^2+(\Gamma_n/2)^2} \;,
\label{LambEff}
\end{eqnarray} 
in full agreement with the previous work of Ref. \cite{VarguetFano:19}.

\subsubsection{Continuous model}
Another expression can be obtained using the Kramers-Kronig relations. From $\mathbf{G}_{scatt}=\mathbf{G}_{tot}-\mathbf{G}_0$, and considering a Drude metal $\epsilon_m(\omega)=\epsilon_\infty-\omega_p^2/(\omega^2+i\Gamma_p\omega)$, it immediately follows that 
\begin{eqnarray}
\mathbf{G}_{scatt}(\omega) \underset{\omega \rightarrow +\infty}{\sim}\mathbf{G}_{\infty}(\omega)-\mathbf{G}_0(\omega) \;,
\end{eqnarray}
where $\mathbf{G}_{\infty}$ is the Green tensor of a homogeneous medium of dielectric constant $\epsilon_\infty$. In the Drude model, $\epsilon_\infty=1$ so that $\mathbf{G}_{scatt}(\omega)$ tends to zero in the high frequency limit for air background ($\epsilon_b=1$) and satisfies the Kramers-Kronig relations \cite{Fleischhauer:2011}
\begin{eqnarray}
\frac{\omega_0^2}{c^2} Re [G_{ij}(\omega_0)] =\frac{1}{\pi} {\mathbb P} \int_{-\infty}^{+\infty} d\omega \frac{\omega^2}{c^2} \frac{Im[G_{ij}(\omega)] }{\omega-\omega_0} \;,
\end{eqnarray}
so that the Lamb shift becomes \cite{Wylie-Sipe:1985b,Fleischhauer:2011}
\begin{eqnarray}
&&\Delta\omega_{jj}  (\omega_0)=\frac{1}{\pi\hbar \epsilon_0} {\mathbb P}\int_{0}^{+\infty}d\omega \frac{\omega ^2}{c^2} \frac{Im G_{jj}(\omega)}{\omega-\omega_0} \\
\nonumber
&&=\frac{\omega_0^2}{\hbar \epsilon_0 c^2}Re [G_{jj}(\omega_0)]-\frac{1}{\pi\hbar \epsilon_0} {\mathbb P}\int_{-\infty}^{0}d\omega \frac{\omega ^2}{c^2} \frac{Im G_{jj}(\omega)}{\omega-\omega_0} \\
\nonumber
&&=\frac{\omega_0^2}{\hbar \epsilon_0 c^2}Re [G_{jj}(\omega_0)]-\frac{1}{\pi\hbar \epsilon_0} {\mathbb P}\int_0^{+\infty}d\omega \frac{\omega ^2}{c^2} \frac{Im G_{jj}(\omega)}{\omega+\omega_0}
\end{eqnarray}
where we used $G_{jj}(-\omega)=G_{jj}^\star(\omega)$. The first term involving the real part of the Green tensor is equivalent to the classical Lamb shift deduced from the driven dipole model \cite{Metiu:1984,Girard-Martin-Dereux:1995b} so that the second term can be considered as a quantum correction (see also appendix \ref{sectSupp:Classic}). It can be recast as proposed by Tian {\it et al} \cite{Tian-Zhao:19} to obtain 
\begin{eqnarray}
\label{eq:LambCorr}
\Delta\omega_{jj}  (\omega_0)&=&\frac{\omega_0^2}{\hbar \epsilon_0 c^2}Re [G_{jj}(\omega_0)]-\frac{1}{2}\Delta\omega_{jj}(0)\\
\nonumber
&&\hspace{1cm}+\omega_0 \int_0^{\infty}d\omega \frac{\omega ^2}{c^2} \frac{Im G_{jj}(\omega)}{\omega(\omega+\omega_0)}\\
\Delta\omega_{jj}(0)&=&\frac{\omega_0^2}{\hbar \epsilon_0c^2}Re [G_{jj}(0)] \\
\nonumber
\Delta\omega_\perp(0)&&\simeq \frac{d^2}{4\pi \hbar \epsilon_0\epsilon_b R^3}\sum_{n=1}^{N}\frac{(n+1)^2}{(1+h/R)^{2n+4}} \\
\nonumber
\Delta\omega_{//}(0)&&\simeq \frac{d^2}{8\pi \hbar \epsilon_0\epsilon_b R^3}\sum_{n=1}^{N}\frac{n(n+1)}{(1+h/R)^{2n+4}}\;,
\end{eqnarray}
where we approximate the Green tensor by its quasi-static limit at low frequency \cite{Zhao:18}. 
\begin{figure}[h]
	\includegraphics[width=0.47\textwidth]{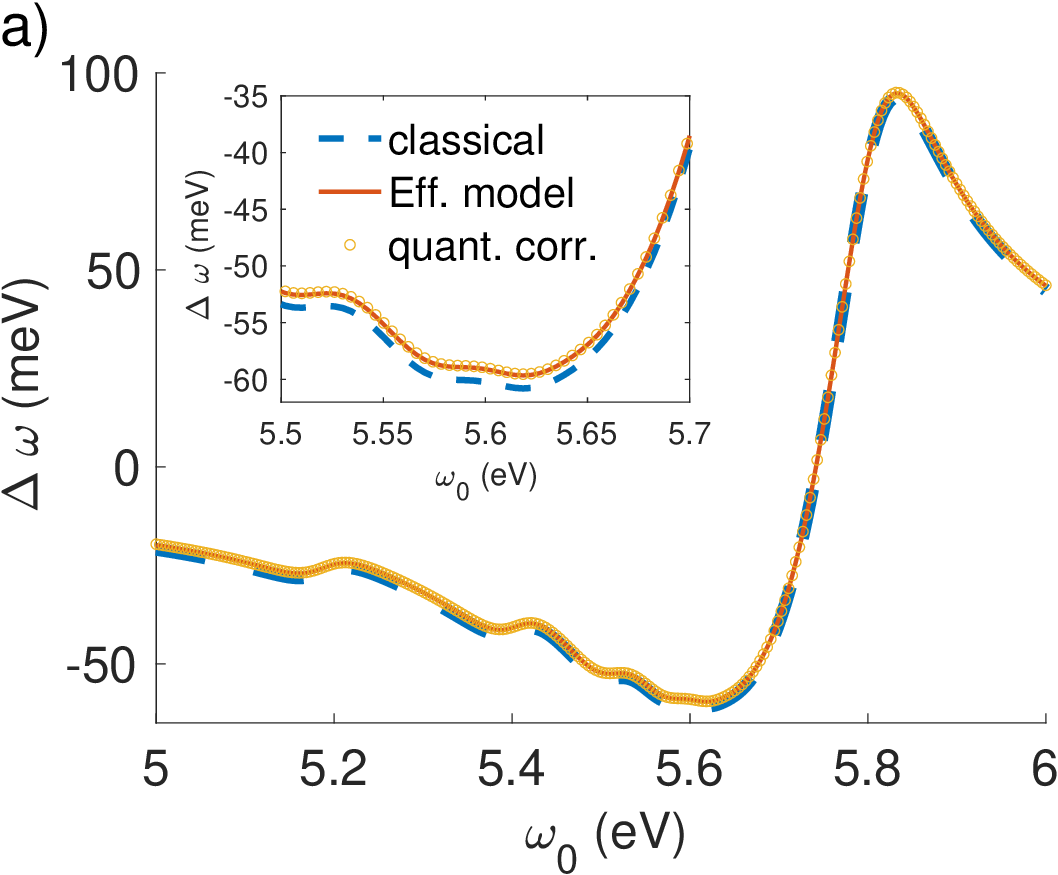}\\
		\includegraphics[width=0.47\textwidth]{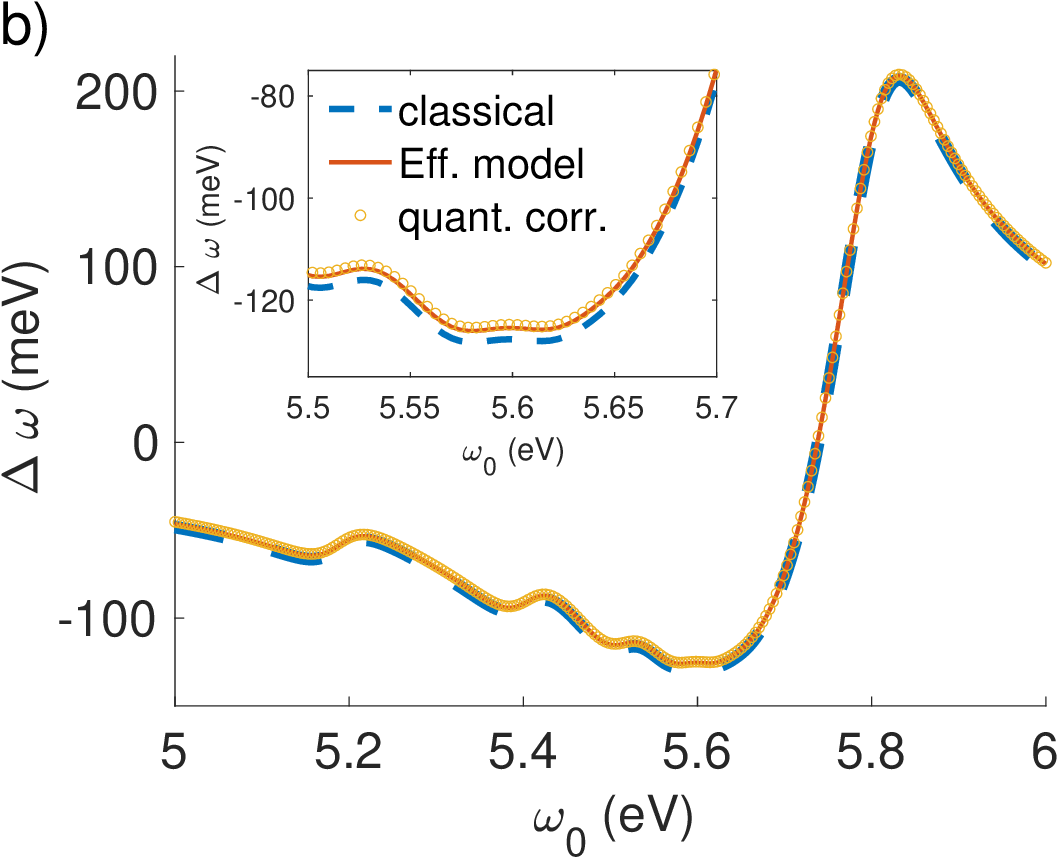}\\
	\caption{Lamb shift calculated as a function of the emission angular frequency $\omega_0$ for a dipolar emitter parallel (a) or perpendicular (b) to the particle surface. The QE is at 2 nm from a 30 nm MNP (R=15 nm), in air ($\epsilon_b=1$).  The parameter of the metal Drude constant are $\epsilon_\infty=1$, $\omega_p=\SI{1.26e16}{\radian \per \second}$ and $\Gamma_p=\SI{1.41e14}{\radian \per \second}$, mimicking gold behaviour in the visible range. "Classical" refers to the first tem of Eq. (\ref{eq:LambCorr}), "quant. corr." corresponds to the full Eq. (\ref{eq:LambCorr}) (including the quantum correction to the classical expression), "Eff. model" is the Lamb shift obtained from the effective model, Eq. (\ref{LambEff}). 
\label{Fig:LambAu}}
\end{figure}
We present in Fig. \ref{Fig:LambAu} the Lamb shift calculated considering the effective model (eq. (\ref{LambEff})), the classical contribution and the classical contribution plus the quantum correction (eq. (\ref{eq:LambCorr})). We observe an excellent agreement between the effective model and the continuous model when the quantum correction is taken into account. However, the quantum correction remains small in this configuration (below  5\%).

\begin{figure}[h]
	\includegraphics[width=0.47\textwidth]{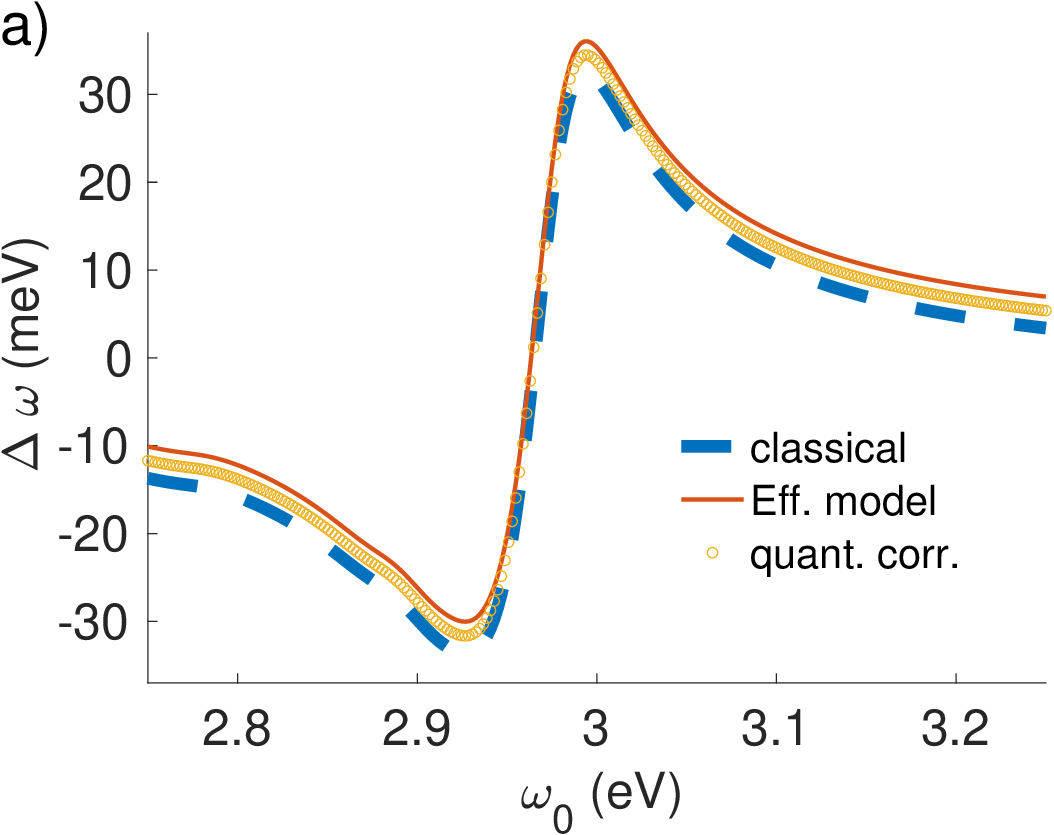}\\
	\includegraphics[width=0.47\textwidth]{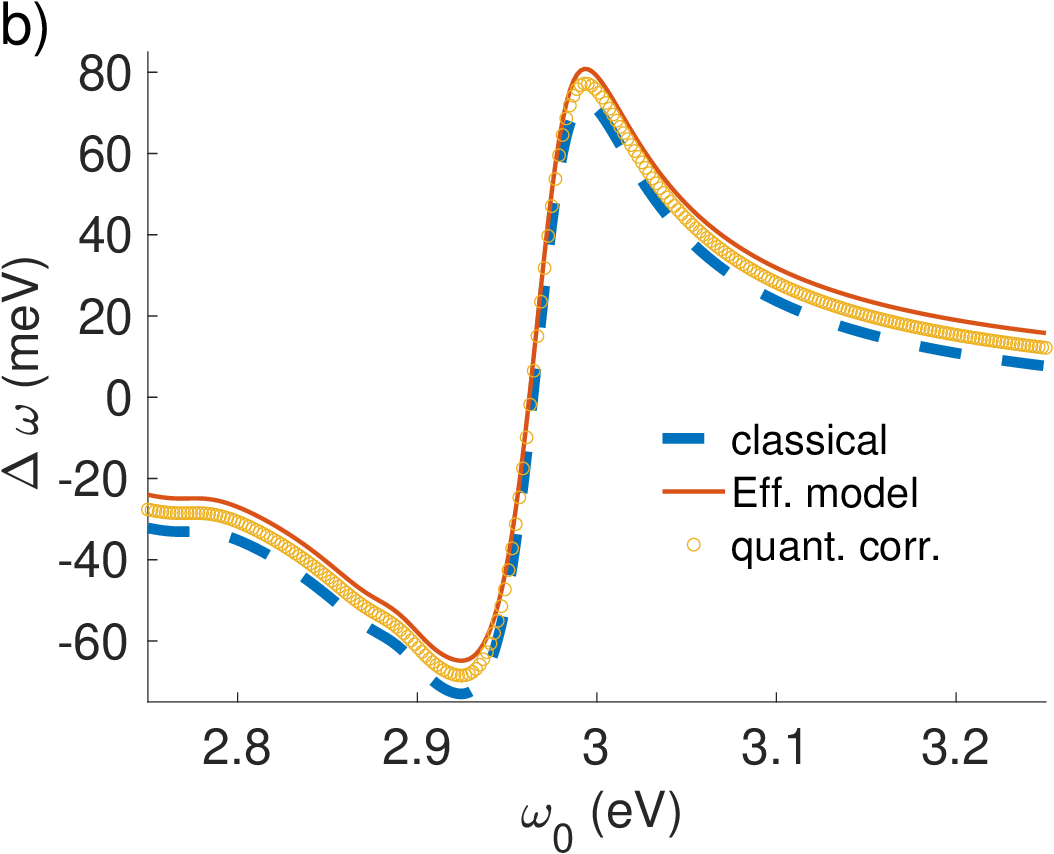}
	\caption{Same as Fig. \ref{Fig:LambAu} but for a silver MNP. $\epsilon_\infty=6$, $\omega_p=\SI{1.20e16}{\radian \per \second}$ and $\Gamma_p=\SI{7.74e13}{\radian \per \second}$.
\label{Fig:LambAg}}
\end{figure}

One generally modifies the Drude model to include interband transitions and better reproduce experimental data. This can be done adding a constant, for instance $\epsilon_\infty=6$ for silver. We observe in Fig. \ref{Fig:LambAg} a small discrepancy between the Lamb shift calculated considering the continuous model (eq. \ref{eq:LambCorr}) and the effective model (eq. \ref{LambEff}). We attribute this difference to the fact that $\mathbf{G}_{scatt}$ does not satisfy the Kramers-Kronig condition for $\epsilon_b \ne \epsilon_\infty$. We check numerically that the agreement is recovered for $\epsilon_b =\epsilon_\infty$ (not shown) and expression (\ref{eq:LambCorr}) is the Lamb shift compared to a homogeneous medium of dielectric constant $\epsilon_\infty$ instead of $\epsilon_b$. Stated differently, since the derivation of eq. (\ref{LambEff}) does not involve the Kramers-Kronig relations, it holds for any configuration and corresponds to the Lamb shift compared to the background medium. 

Since the quantum correction remains small in all the cases discussed above, the classical Lamb shift can be satisfactorily used to compute the emission spectrum in the continuous model, simplifiying the computing task. However, for a multi-emitter configuration, we will demonstrate in section \ref{sect:SpecNe} that the Lamb shift is proportionnal to the number of emitters when all emitters are at the same location, so that the quantum correction cannot be neglected anymore. 

Since the effective models holds, we can estimate the quantum correction without numerically computing the integral as 
\begin{eqnarray}
\Delta \omega_c^{(j)} &=&-\frac{1}{\pi\hbar \epsilon_0} {\mathbb P}\int_{-\infty}^0d\omega \frac{\omega ^2}{c^2} \frac{Im G_{jj}(\omega)}{\omega-\omega_0}\\
\nonumber
&=&\sum_{n=1}^{N}\frac{[g_n^{(j)}]^2(\omega_0-\omega_n)}{(\omega_0-\omega_n)^2+(\Gamma_n/2)^2}-\frac{\omega_0^2}{\hbar \epsilon_0c^2}Re [G_{jj}(\omega_0)]
\end{eqnarray}
that is valid independantly of the value of $\epsilon_b$.

\subsection{Dipole-dipole shift}
Simarly, for $i\ne j$, the dipole-dipole shift is expressed using the effective model \cite{Varguet-GCF:19}
\begin{eqnarray}
\Delta\omega_{ij}(\omega_0)&=&\frac{1}{\pi\hbar \epsilon_0} {\mathbb P}\int_{0}^{+\infty}d\omega \frac{\omega ^2}{c^2} \frac{Im G_{ij}(\omega)}{\omega-\omega_0}\\
\nonumber
&=&\sum_{n=1}^{N}\frac{(\omega_0-\omega_n)}{(\omega_0-\omega_n)^2+(\Gamma_n/2)^2}g_n^{(i)}g_n^{(j)} \mu_n^{i,j}(\omega_n) \;.
\label{Dip-DipEff}
\end{eqnarray} 

In the context of the continuous model, we can write 
\begin{eqnarray}
\nonumber
\Delta\omega_{ij}  (\omega_0)&=&\frac{\omega_0^2}{\hbar \epsilon_0c^2}Re [G_{ij}(\omega_0)] \\
\nonumber
&&\hspace{2cm} -\frac{1}{\pi\hbar \epsilon_0} {\mathbb P}\int_{-\infty}^0d\omega \frac{\omega ^2}{c^2} \frac{Im G_{ij}(\omega)}{\omega-\omega_0} \\
&=&\frac{\omega_0^2}{\hbar \epsilon_0c^2}Re [G_{ij}(\omega_0)]+\Delta \omega_c^{(ij)} \;,
\label{Dip-DipCorr}
\end{eqnarray} 
where $\Delta \omega_c^{(ij)} $ is the quantum correction to the classical dipole-dipole shift that is easily calculated as the difference between the effective model and classical model expressions of the dipole-dipole shift. We estimate the quantum correction (including Lamb shift and dipole-dipole quantum corrections) for each emitter to $\Delta \omega_c=\vert\sum_{j=1}^{N_e}\Delta \omega_c^{(1j)}\vert=22$ meV for $N_e=50$ emitters in the ring configuration. It increases to $\Delta \omega_c=219$ meV for $N_e=500$.
\subsection{Effect of quantum corrections on the emission spectrum}
\subsubsection{Single emitter configuration}
\begin{figure}[h]
	\includegraphics[width=0.47\textwidth]{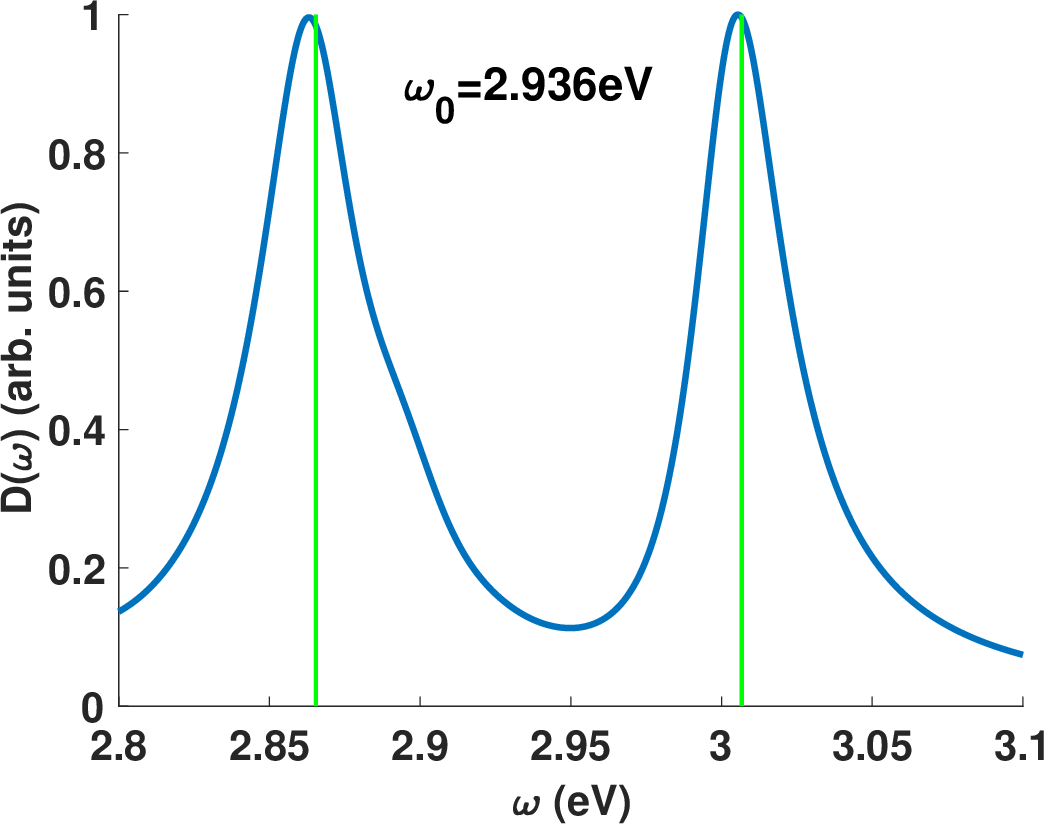}\\
	\includegraphics[width=0.47\textwidth]{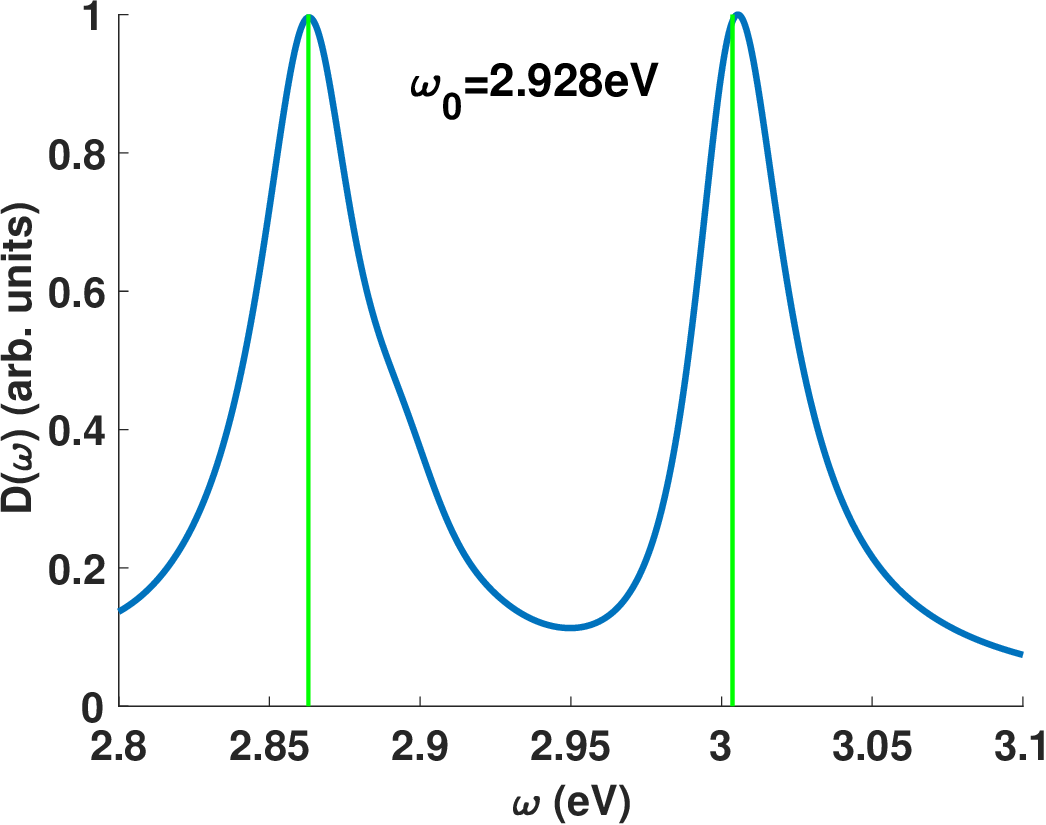}
		\caption{Emission spectrum for a single QE coupled to a silver MNP using the continuous model without (a) or with (b) the quantum correction. Vertical lines indicate the energies of the two main dressed states contributing to the coupling. The emitter is orthogonal to the MNP. Other parameters are the same as in Fig. \ref{Fig:LambAg}. The emission frequency leading to the strong coupling is indicated on the figures. The quantum correction is $\Delta \omega_c(\omega_0)=-7.5$ meV. 
\label{Fig:Spec1}}
\end{figure}

For a single emitter coupled to a MNP, the emission spectrum is expressed in the continuous model, see Eq. (\ref{eq:SpecCont}) \cite{Hakami-Wang-Zubairy:2014}
\begin{eqnarray}
D(\omega)&=&\frac{\gamma_0}{2\pi} \left \vert \frac{1}{ \omega-\omega_0+\Delta \omega_c(\omega)+i\frac{\gamma_0}{2}+\frac{k_0^2 G_{11}(\omega)}{\hbar \epsilon_0}}\right \vert ^2 \,.
\hspace{0.5cm}
\end{eqnarray}
We observe that the quantum correction does not significantly change over the considered spectral range so that we can safely replace $\Delta \omega_c(\omega)\approx \Delta \omega_c(\omega_0)$ and 

\begin{eqnarray}
D(\omega)&=&\frac{\gamma_0}{2\pi} \left \vert \frac{1}{ \omega-\omega_0+\Delta \omega_c(\omega_0)+i\frac{\gamma_0}{2}+\frac{k_0^2 G_{11}(\omega)}{\hbar \epsilon_0}} \right \vert ^2 \,.
\hspace{0.5cm}
\end{eqnarray}
The quantum correction clearly introduces a small resonance shift in the emission spectrum. This correction can be neglected as shown in Fig. \ref{Fig:Spec1}. 

\subsubsection{Multi-emitter configuration}
\label{sect:SpecNe}
\begin{figure}[h]
	\includegraphics[width=0.47\textwidth]{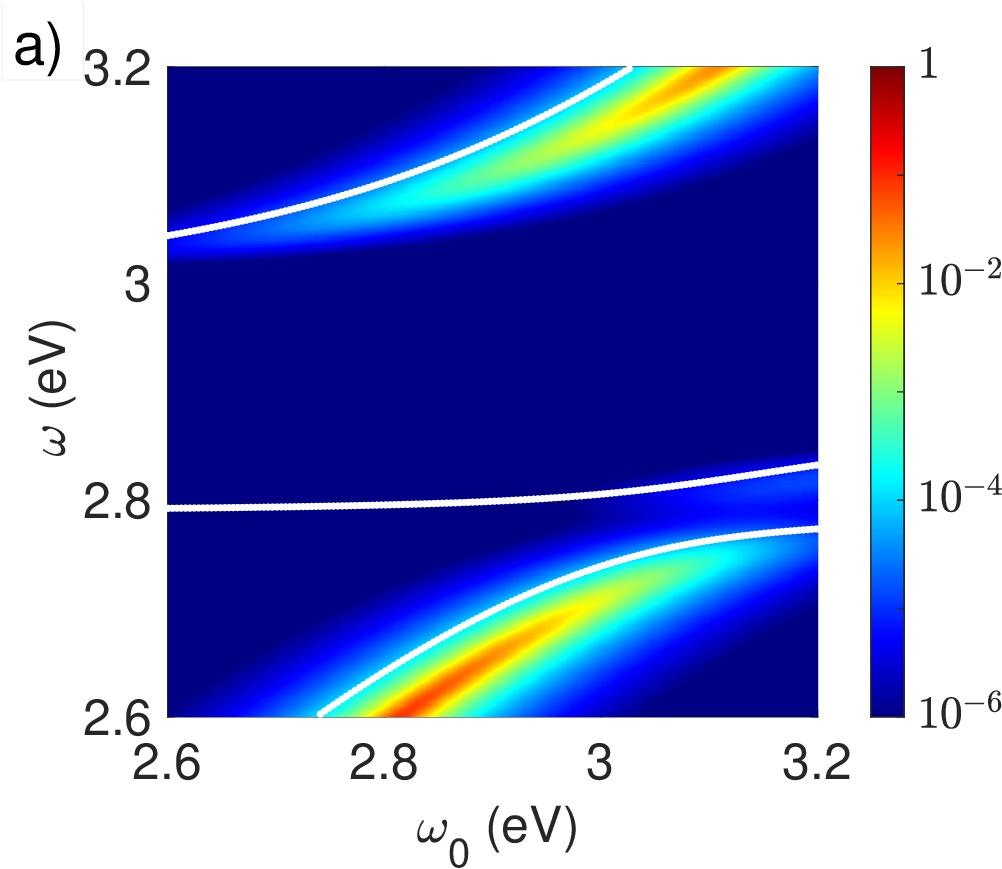}\\
	\includegraphics[width=0.47\textwidth]{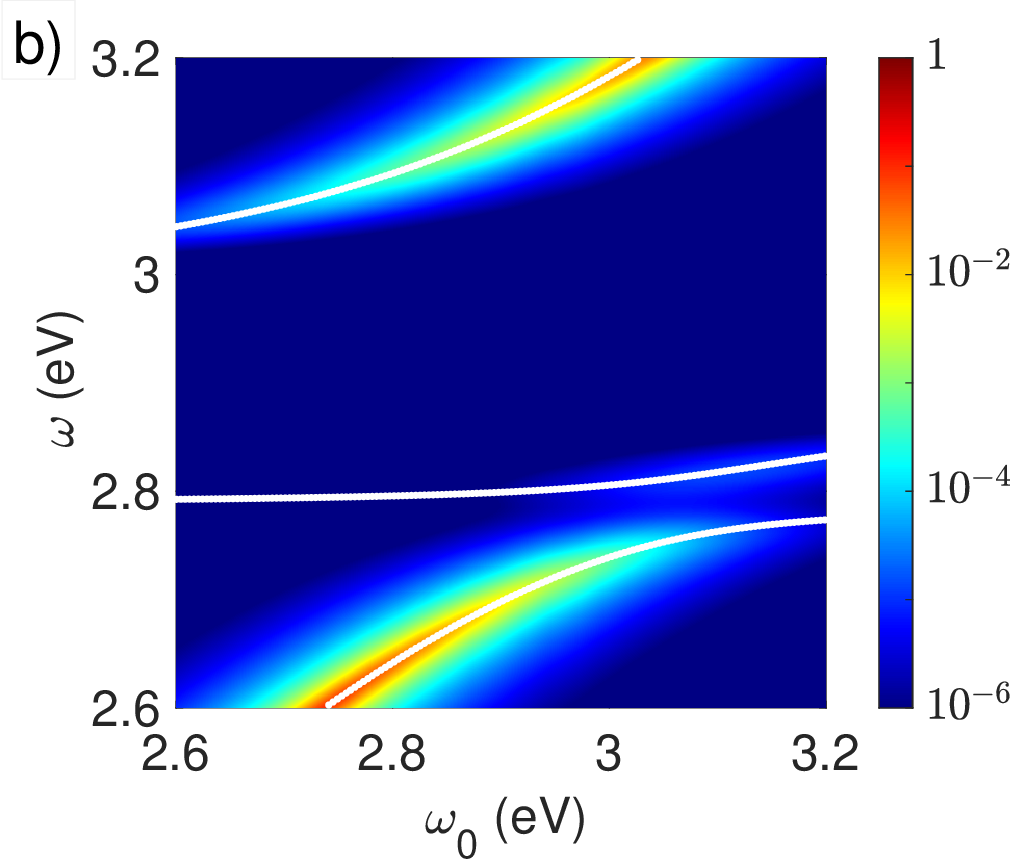}
	\caption{Emission spectrum for $N_e=25$ QEs coupled to a silver MNP using the continuous model without (a) or with (b) the quantum correction. Solid white lines indicate the angular frequencies $\Omega_m$ of the dressed states that contribute mainly to the strong coupling.
\label{Fig:Spec10}}
\end{figure}

For several emitters, an additionnal dipole-dipole shift occurs. It decreases fastly as a function of the distance between two emitters so that we restrict our study to the ideal configuration to demonstrate the role of quantum corrections. In this case, all the elements of the matrix ${\mathbf M}$ are identical $M_{ij}=M_{11} \;, \forall i,j$ and the emission spectrum takes the simple form (see Eq. (\ref{eqSupp:SpecContIdeal})) 
 \begin{eqnarray}
D(\omega)&\propto &\left \vert \frac{1}{ \omega-\omega_0+N_e \Delta \omega_c+i\frac{\gamma_0}{2}+\frac{N_e k_0^2}{\hbar \epsilon_0} G_{11}(\omega) }\right \vert ^2\,,
\hspace{0.5cm}
\end{eqnarray}
so that the cumulative effect of quantum corrections $N_e \Delta \omega_c(\omega_0)$ is now clearly apparent and cannot be ignored for a large number of QEs. A significant spectral shift is observed neglecting  the quantum corrections, as shown {\it e.g} in Fig. \ref{Fig:Spec10} for only 25 emitters. 

\begin{figure}[h]
	\includegraphics[width=0.47\textwidth]{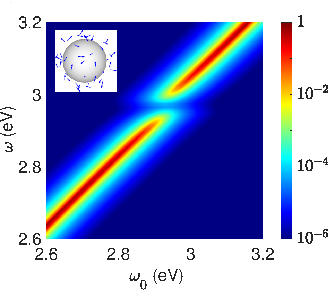}
	\caption{Emission spectrum for $N_e=50$ randomly oriented emitters spread around the MNP (see inset).}
\label{Fig:SpecRand50}
\end{figure}

As a last configuration, we consider 50 emitters randomly oriented all around the particle, see Fig. \ref{Fig:SpecRand50}. The Rabi splitting remains low (about 100 meV) compared to the ideal and ring configurations. This originates from different frequency shifts for each emitter, notably because of the quantum corrections. Indeed, we compute Van der Waals frequency shifts ranging from -15 meV to +15 meV and quantum corrections ranging from -35 meV to 5 meV. This leads to a different strong coupling gap frequency for each emitter, hence a lower total Rabi splitting.

\section{Conclusion}
We have investigated the collective strong coupling regime in a plasmonic nanocavity. We observe that the Rabi splitting is proportionnal to $\sqrt{N_e}$ if all the emitters are located at the same position, but it strongly deviates from this behaviour for a homogeneous ring configuration around the particle, severely decreasing the collective strong coupling behaviour. We derive a continuous approach and an effective model, bringing a clear physical understanding of the collective coupling process, emphasizing notably the role of LSP modes. In addition, the comparison between the two equivalent approaches leads to a full understanding and estimation of quantum corrections to the classical expression for the Lamb shift. We also identify a collective Lamb shift proportionnal to the number of emitters when located at the same position. 
The effective model intrinsically includes quantum corrections whereas the continuous model necessitates their carefull estimation. Quantum corrections affect the dipole-dipole coupling and introduce a frequency shift. When they are neglected, the continous model simplifies to the classical approach.
A better understanding of the collective behaviour of multiple quantum emitters strongly coupled to a plasmonics nanocavity is crucial for optimizing the hybrid system, paving the way towards the generation of non classical plasmon states in analogy to cQED devices \cite{Faraon-Vukovic:08,Iliopoulos-Paspalakis:19} or controling chemical reactions \cite{Shalabney-Ebbesen:2015,Kato:18,Herrera-Owrutsky:20}.

\begin{acknowledgments}
%The authors thank A. Diaz-Valles for complementary numerical simulations and helpful discussions. 
This work is supported by the French "Investissements d'Avenir" program, through the project ISITE-BFC IQUINS (contract ANR-15-IDEX-03 ) and EUR-EIPHI contract (17-EURE-0002). We acknowledge support from the European Union’s Horizon 2020 research and innovation program under the Marie Sklodowska-Curie grant agreement No. 765075 (LIMQUET).
\end{acknowledgments}

The data that support the findings of this study are available from the corresponding author upon reasonable request.

\newpage
\onecolumngrid
\appendix
\section{Emission spectrum in the continuous model}
\label{sectSupp:ContSpec}
The dynamics of the wavefunction (Eq. (\ref{wavefun})) is governed by the Schr\"odinger equation (Eq. (\ref{eq:schrodinger})) with the Hamiltonian (\ref{hamil}) so that 
\begin{subequations}
\begin{eqnarray}
{\dot{\tilde{C}}_{e,\mathbf{0}}^{(l)}(t)}&=&-\frac{1}{\sqrt{\hbar\pi\varepsilon_0}}\int_0^{+\infty}\!\!\!\!\!\mathrm{d}\omega\int \mathrm{d}^3r\; \frac{\omega^2}{c^2}\sqrt{\varepsilon_I(\mathbf{r},\omega)}\mathbf{d}^{(l)}\cdot\bar{\bar{\mathbf{G}}}(\mathbf{r}_l,\mathbf{r},\omega)\cdot\mathbf{C}_g (\mathbf{r},\omega,t)\e^{i(\omega_0-\omega)t}-\frac{\gamma_0}{2}{\tilde{C}_{e,\mathbf{0}}(t)}\label{Ce_dyn_annexe_chap4},\\
\dot{\mathbf{C}}_g(\mathbf{r},\omega,t)&=&\frac{1}{\sqrt{\hbar\pi\varepsilon_0}}\frac{\omega^2}{c^2}\sqrt{\varepsilon_I(\mathbf{r},\omega)}\sum_{l=1}^{N_e}\bar{\bar{\mathbf{G}}}^*(\mathbf{r},\mathbf{r}_l,\omega)\cdot\left(\mathbf{d}^{(l)}\right)^* \;{\tilde{C}_{e,\mathbf{0}}^{(l)}(t)}\e^{i(\omega-\omega_0)t}\label{Cg_dyn_annexe_chap4}{,}
\end{eqnarray}
\end{subequations}
{with $\tilde{C}_{e,\mathbf{0}}^{(l)}(t)=C_{e,\mathbf{0}}^{(l)}(t)\,\e^{i\omega_0t}$.} We first define the spectral density for each emitter:
\begin{subequations}
\begin{eqnarray}
D_l(\omega)&=&\frac{\gamma_0^{(l)}}{2\pi}\left\vert \int_0^{+\infty}\!\!\!\!\!{\tilde{C}_{e,\mathbf{0}}^{(l)}(t)}\e^{i(\omega-\omega_0) t}\mathrm{d}t\right\vert^2,\label{def_D_annexe_chap4}
\\
&=&\frac{\gamma_0}{2\pi}\left\vert \mathcal{FT}\left[{\tilde{C}_{e,\mathbf{0}}^{(l)}(t)}\right](\omega)\right\vert^2,
\label{def_D_TF_annexe_chap4}
\end{eqnarray}
\end{subequations}
where $FT$ denotes the Fourier transform. 

Formally integrating equation (\ref{Cg_dyn_annexe_chap4}), Eq (\ref{Ce_dyn_annexe_chap4}) is expressed as 
\begin{subequations}
\begin{eqnarray}
{\dot{\tilde{C}}_{e,\mathbf{0}}^{(l)}(t)}=&-\frac{1}{\hbar\pi\varepsilon_0}\int_0^{+\infty}\!\!\!\!\!\mathrm{d}\omega\;\frac{\omega^2}{c^2}\e^{i(\omega_0-\omega)t}\sum_{j=1}^{N_e}\Bigg[\int_0^t \mathrm{d}t'\; {\tilde{C}_{e,\mathbf{0}}^{(j)}(t')}\e^{i(\omega-{\omega_0})t'}\nonumber\\
&\times\int \mathrm{d}^3r\;\frac{\omega^2}{c^2}\varepsilon_I(\mathbf{r},\omega)\mathbf{d}^{(l)}\cdot\bar{\bar{\mathbf{G}}}(\mathbf{r}_l,\mathbf{r},\omega) \bar{\bar{\mathbf{G}}}^*(\mathbf{r},\mathbf{r}_j,\omega)\cdot\left(\mathbf{d}^{(j)}\right)^*\Bigg]-\frac{\gamma_0}{2}{\tilde{C}_{e,\mathbf{0}}^{(l)}(t)},\label{Ce2_dyn_annexe_chap4}\\
{\dot{\tilde{C}}_{e,\mathbf{0}}^{(l)}(t)}=&\int_0^t \mathrm{d}t'\; \sum_{j=1}^{N_e} \mathcal{K}_{lj}(t-t'){\tilde{C}_{e,\mathbf{0}}^{(j)}(t')}-\frac{\gamma_0}{2}{\tilde{C}_{e,\mathbf{0}}^{(l)}(t)}\label{Ce3_dyn_annexe_chap4},
\end{eqnarray}
\end{subequations}
with
\begin{eqnarray}
\mathcal{K}_{lj}(t-t')=-\frac{1}{\hbar\pi\varepsilon_0}\int_0^{+\infty}\!\!\!\!\!\mathrm{d}\omega\; \frac{\omega^2}{c^2} \e^{-i(\omega-\omega_0)(t-t')}\;\mathbf{d}^{(l)}\cdot\mathfrak{Im}\left[\bar{\bar{\mathbf{G}}}(\mathbf{r}_l,\mathbf{r}_j,\omega)\right]\cdot\left(\mathbf{d}^{(j)}\right)^*\,,
\end{eqnarray}
where we used the property 
\begin{eqnarray}
\int \mathrm{d}^3r\;\frac{\omega^2}{c^2}\varepsilon_I(\mathbf{r},\omega)\bar{\bar{\mathbf{G}}}(\mathbf{r}_l,\mathbf{r},\omega)\bar{\bar{\mathbf{G}}}^*(\mathbf{r},\mathbf{r}_j,\omega)=\mathfrak{Im}\left[\bar{\bar{\mathbf{G}}}(\mathbf{r}_l,\mathbf{r}_j,\omega)\right],
\end{eqnarray} 
Finally, 
\begin{eqnarray}
-{\tilde{C}_{e,\mathbf{0}}^{(l)}(0)}-i(\omega-\omega_0)\mathcal{FT}\left[{\tilde{C}_{e,\mathbf{0}}(t)}\right](\omega)=\sum_{j=1}^{N_e}\mathcal{FT}\left[\mathcal{K}_{lj}(t)\right](\omega)\mathcal{FT}\left[{\tilde{C}_{e,\mathbf{0}}^{(j)}(t)}\right](\omega)-\frac{\gamma_0}{2}\mathcal{FT}\left[{\tilde{C}_{e,\mathbf{0}}^{(l)}(t)}\right](\omega),\label{TF_annexe_chap4}
\end{eqnarray}
with  ${\tilde{C}_{e,\mathbf{0}}^{(l)}(0)}$ the initial condition on the state $\vert e^{(l)},\mathbf{0}\rangle$.  The Fourier transform of $\mathcal{K}_{lj}(t)$ is expressed as 
\begin{subequations}
\begin{eqnarray}
M_{lj}(\omega)&=&\frac{{i}}{\hbar \pi \epsilon_0}\int_0^{\infty} d\omega' \frac{\omega'^2}{c^2} ImG_{lj}(\omega') \int_0^\infty dt e^{-i(\omega'-\omega)t} \\
\label{eqSupp:Lamb}
&=&\frac{{i}\omega^2}{\hbar \epsilon_0c^2}Im[G_{lj}(\omega)] +\frac{{1}}{\hbar \pi \epsilon_0} {\mathbb P} \int_0^{\infty} d\omega' \frac{\omega'^2}{c^2} \frac{Im[G_{lj}(\omega')]}{\omega'-\omega}\\
G_{lj}(\omega)&=&\mathbf{d}^{(l)} \cdot \mathbf{G}_{scatt}(\mathbf{r}_l,\mathbf{r}_j,\omega)\cdot \left(\mathbf{d}^{(j)}\right)^\star\,,
\end{eqnarray}
\end{subequations}
where we have used the property (see also \cite{Hakami-Wang-Zubairy:2014} for the single emitter case $N_e=1$)
\begin{eqnarray}
 \int_0^\infty dt e^{-i(\omega'-\omega)t}=\pi\delta(\omega'-\omega)-i{\mathbb P}\left(\frac{1}{\omega'-\omega}\right)
\end{eqnarray}

In order to solve the system of equation (\ref{TF_annexe_chap4}), we define the column vector
\begin{eqnarray}
{\tilde{C}_{e,\mathbf{0}}(t)}=\begin{bmatrix}
{\tilde{C}_{e,\mathbf{0}}^{(1)}(t)}\\
{\tilde{C}_{e,\mathbf{0}}^{(2)}(t)}\\
{\tilde{C}_{e,\mathbf{0}}^{(3)}(t)}\\
\vdots\\
{\tilde{C}_{e,\mathbf{0}}^{(N_e)}(t)}\\\end{bmatrix},\label{Vec_Ce_annexe_chap4}
\end{eqnarray}
so that %
\begin{eqnarray}
-{\tilde{C}_{e,\mathbf{0}}(0)}+i(\omega_0-\omega)\mathcal{FT}\left[{\tilde{C}_{e,\mathbf{0}}(t)}\right](\omega)={i}{\mathbf{M}(\omega)}\;\mathcal{FT}\left[{\tilde{C}_{e,\mathbf{0}}(t)}\right](\omega)-\frac{\gamma_0}{2}\mathcal{FT}\left[{\tilde{C}_{e,\mathbf{0}}(t)}\right](\omega) \;,
\end{eqnarray}
and  
\begin{eqnarray}
\mathcal{FT}\left[{\tilde{C}_{e,\mathbf{0}}(t)}\right](\omega)={\frac{1}{i}\left[\left(\omega_0-\omega-i\frac{\gamma_0}{2}\right)\mathbf{I}-\mathbf{M}(\omega)\right]^{-1}\tilde{C}_{e,\mathbf{0}}(0)} \,.
\label{blabla_annexe_chap4}
\end{eqnarray}
We obtain the spectral density of emitter $l$:
\begin{eqnarray}
D_l(\omega)=\frac{\gamma_0}{2\pi}\left\vert\left\{\left[\left(\omega-\omega_0)+i\frac{\gamma_0}{2}\right)\mathbf{I}+\mathbf{M}(\omega)\right]^{-1}{\tilde{C}_{e,\mathbf{0}}(0)}\right\}_l\right\vert^2 \;,
\label{D_Ne_annexe4}
\end{eqnarray}
and the spectral density of the complete hybrid system is expressed as
\begin{eqnarray}
D(\omega)=\frac{\gamma_0}{2\pi}\left\vert\int_0^{+\infty}\!\!\!C_{\alpha}(t)\e^{i\omega t}\mathrm{d}t\right\vert^2,
\end{eqnarray}
where $C_{\alpha}(t)$ is the probability amplitude of the state
\begin{eqnarray}
\vert\alpha\rangle=\frac{1}{\sqrt{\sum_{l=1}^{N_e}\left\vert a_l\right\vert^2}}\sum_{l=1}^{N_e}a_{l}\vert e^{(l)},\mathbf{0}\rangle.
\end{eqnarray}
The projection of state  $\vert\alpha\rangle$ on the wavefunction  {(\ref{wavefun})} leads to 
\begin{eqnarray}
{D(\omega)=\frac{\gamma_0}{2\pi}\frac{1}{\sum_{l=1}^{N_e}\left\vert a_l\right\vert^2}\left\vert\sum_{l=1}^{N_e}a_l^*\left\{\left[\left(\omega-\omega_0)+i\frac{\gamma_0}{2}\right)\mathbf{I}_{N_e}+M(\omega)\right]^{-1}\tilde{C}_{e,\mathbf{0}}(0)\right\}_l\right\vert^2.}\label{D_globale_annexe_chap4}
\end{eqnarray} 
and for the initial coherent superposition $\ket{\psi(0)}=\frac{1}{\sqrt{N_e}}\sum_{l=1}^{N_e}\ket{e^{(l)},\vac}$, 
the spectral density simplifies to Eq. (\ref{eq:SpecCont}) in the main text.

\section{Configuration with all QEs at the same position}
For all QEs located at the same position, and considering the initial condition $\vert\psi(0)\rangle=\vert B\rangle$, all the element of the matrix $\mathbf{M}$ are identical, $M_{ij}(\omega)=M_{11}(\omega)=k_0^2/(\hbar \epsilon_0)G_{11}(\omega)+\Delta \omega_c$ and $C^{(l)}_{e,\varnothing}(0)=1/\sqrt{N_e}$ $\forall\,l$, so that the spectral density is expressed as
\begin{subequations}
\begin{eqnarray}
\label{eqSupp:SpecContIdeal}
D(\omega)&=&\frac{\gamma_0}{2\pi N_e^2} \left \vert \sum_{k=1}^{N_e}\sum_{l=1}^{N_e}\left\{ \left[ \left( (\omega-\omega_0)+i \frac{\gamma_0}{2}\right)\mathbb{\mathbf{I}}+M_{11}(\omega)\mathbf{P}\right]^{-1}\right\}_{kl} \right \vert^2  \,, \\
\mathbf{P}&=&
\begin{bmatrix}
1 &1 &\cdots & 1\\
1 &1 &\cdots & 1\\
\vdots &\vdots &  \ddots & \vdots\\
1 & 1 &\cdots & 1 \end{bmatrix} \;.
\end{eqnarray}
\end{subequations}
One can simplify the expression of $D(\omega)$ using the property 
\begin{eqnarray}
 \left[ a\mathbb{\mathbf{I}}+b\mathbf{P}\right]^{-1}=\frac{1}{a}\mathbb{\mathbf{I}}-\frac{b}{a(a+N_e b)}\mathbf{P} \;,
\end{eqnarray}
with $a=\omega-\omega_0+i \gamma_0/2$ and $b=M_{11}(\omega)$ so that 
\begin{subequations}
\begin{eqnarray}
D(\omega)&=&\frac{\gamma_0}{2\pi N_e^2} \frac{1}{(\omega-\omega_0)^2+\gamma_0/2)^2}\left \vert \sum_{k=1}^{N_e}\sum_{l=1}^{N_e}\left\{ \mathbb{\mathbf{I}}-\frac{M_{11}(\omega)}{(\omega-\omega_0)+i \frac{\gamma_0}{2}+N_eM_{11}(\omega)}\mathbf{P}\right\}_{kl} \right \vert^2 \,, \\
&=&\frac{\gamma_0}{2\pi N_e} \frac{1}{(\omega-\omega_0)^2+\gamma_0/2)^2}\left \vert 1-\frac{N_e M_{11}(\omega)}{(\omega-\omega_0)+i \frac{\gamma_0}{2}+N_e M_{11}(\omega)}\right \vert^2 \,, \\
&=&\frac{\gamma_0}{2\pi N_e} \left \vert \frac{1}{(\omega-\omega_0)+i \frac{\gamma_0}{2}+N_e M_{11}(\omega)}\right \vert^2 \,, \\
&=&\frac{\gamma_0}{2\pi N_e} \left \vert \frac{1}{(\omega-\omega_0+N_e\Delta \omega_c)+i \frac{\gamma_0}{2}+N_e k_0^2/(\hbar \epsilon_0)G_{11}(\omega)}\right \vert^2 \,.
\label{eqSupp:SpecContIdeal}
\end{eqnarray}
\end{subequations}
\section{Classical Drude-Lorentz model}
\label{sectSupp:Classic}
A classical derivation of the collective strong coupling can be  done considering the Drude-Lorentz driven dipole model. Each dipolar emitter is modeled by an oscillating dipole ${\bf d}^{(i)}$. %For comparison purpose, we assume that all the dipoles are normalized, so that it corresponds to all the emitter in their exciting state, defining the classical analogue to the initial coherent superposition discussed in the main text.    
In presence of an electric field ${\bf  E}$, the  dynamics of each emitter obeys the equation \cite{Metiu:1984}
\begin{eqnarray}
-\omega^2{\bf d}^{(i)}-i\omega \gamma_0 {\bf d}^{(i)}  +\omega_0^2 {\bf d}^{(i)} =\frac{e^2}{m}{\bf f}_i \cdot {\bf E}({\bf r}_i) \;,(i=1,\ldots , Ne)\;,
\end{eqnarray}
where $e$ is the elementary charge and $m$ the electron mass, respectively. ${\bf f}$ is the oscillator strength of the i$^{th}$ emitter. 
The electric field scattered at position ${\bf r}_i$, by the dipole ${\bf d}^{(j)} $ is 
\begin{eqnarray}
{\bf E}({\bf r}_i)=\frac{k_0^2}{\epsilon_0}\mathbf{G}_{tot}({\bf r}_i,{\bf r}_j,\omega)\cdot {\bf d}^{(j)} 
\end{eqnarray}
so that the dynamics of the ensemble of oscillators obeys 
\begin{subequations}
\begin{eqnarray}
&&-\omega^2 d^{(i)}  -i\gamma_0 \omega  d^{(i)}   +\omega_0^2 d^{(i)}  =\frac{k_0^2e^2}{m\epsilon_0} \sum_{j=1}^{N_e} {\bf d}^{(i)} \cdot \mathbf{G}_{tot}({\bf r}_i,{\bf r}_j,\omega)\cdot {\bf d}^{(j)}  (i=1,\ldots, N_e) \;,\\
&&[(\omega_0^2-\omega^2)  -i\gamma_0 \omega ]d^{(i)} 
-\frac{e^2}{m}\mu_0\omega^2\sum_{j=1}^{N_e} {\bf d}^{(i)} \cdot \mathbf{G}_{tot}({\bf r}_i,{\bf r}_j,\omega)\cdot {\bf d}^{(j)}  (i=1,\ldots, N_e)\;.
\end{eqnarray}
\end{subequations}
Close to atomic transition $\omega_0$, we approximate $(\omega_0^2-\omega^2)=(\omega_0+\omega)(\omega_0-\omega)\approx 2\omega_0(\omega_0-\omega)$ so that 
\begin{subequations}
\begin{eqnarray}
&&[(\omega-\omega_0) +i\frac{\gamma_0}{2} ]d^{(i)} 
+\frac{e^2}{2m}\mu_0\omega_0\sum_{j=1}^{N_e} {\bf d}^{(i)} \cdot \mathbf{G}_{tot}({\bf r}_i,{\bf r}_j,\omega_0)\cdot {\bf d}^{(j)} =0 \;, (i=1,\ldots, N_e) \;,\\
&&i\frac{\gamma_0}{2}d^{(i)} 
+\frac{e^2}{2m}\mu_0\omega_0\sum_{j=1}^{N_e} {\bf d}^{(i)} \cdot \mathbf{G}_{tot}({\bf r}_i,{\bf r}_j,\omega_0)\cdot {\bf d}^{(j)} =(\omega_0-\omega)d^{(i)} \;, (i=1,\ldots, N_e) \;.
\end{eqnarray}
\end{subequations}
Finally, using the classical radiation reaction linewidth \cite{Novotny-Hecht:2012} $\gamma_0=2e^2\omega_0^2/12\pi\epsilon_0mc^3$, it comes 
\begin{eqnarray}
i\frac{\gamma_0}{2}d^{(i)} 
+\frac{\gamma_0}{2}\frac{6\pi}{k_0}\sum_{j=1}^{N_e} {\bf d}^{(i)} \cdot \mathbf{G}_{tot}({\bf r}_i,{\bf r}_j,\omega_0)\cdot {\bf d}^{(j)} =(\omega_0-\omega)d^{(i)} \;, (i=1,\ldots, N_e)\;,
\end{eqnarray}
so that it is equivalent to finding the eigenvalue of the matrix 
\begin{eqnarray}
\label{eq:classic}
 i\frac{\gamma_0}{2}\left[\mathbf{I}+\frac{6\pi}{k_0}\mathbf{K}\right] \;, \text{with}\;  
 K_{ij}={\bf d}^{(i)} \cdot \mathbf{G}_{tot}({\bf r}_i,{\bf r}_j,\omega_0)\cdot {\bf d}^{(j)}
\end{eqnarray}

At this point, it is useful to compare the quantum and classical model. Remembering the expression of the spectral density obtained within the quantum continuous model (Eq. \ref{eq:SpecCont}) and the expression of the matrix $\mathbf{M}(\omega)$ simplified using Kramers-Kronig relations (Eq. \ref{eq:Mij} and \ref{Dip-DipCorr}) 
\begin{subequations}
\begin{eqnarray}
\label{eq:SpecContbis}
&&D(\omega) \propto \left \vert \sum_{k=1}^{N_e}\sum_{l=1}^{N_e}\left\{ \left[ \left( (\omega-\omega_0)+i\frac{\gamma_0}{2}\right)\mathbb{\mathbf{I}}+\mathbf{M}(\omega)\right]^{-1}\right\}_{kl} \right \vert^2 \,,\\
&&M_{ij}(\omega)=\frac{k_0^2}{\hbar \epsilon_0}{\bf d}^{(i)} \cdot \mathbf{G}_{tot}({\bf r}_i,{\bf r}_j,\omega_0)\cdot {\bf d}^{(j)}+\Delta \omega_c^{(ij)} \\
&& \Delta \omega_c^{(ij)}=-\frac{1}{\hbar \pi \epsilon_0} {\mathbb P} \int_{-\infty}^0 d\omega \frac{\omega^2}{c^2} \frac{Im[G_{ij}(\omega)]}{\omega-\omega_0}
\end{eqnarray}
\end{subequations}
If we neglect the quantum corrections $\Delta \omega_c^{(ij)}$, it reduces to 
\begin{subequations}
\begin{eqnarray}
&&D_{class}(\omega) \propto \left \vert \sum_{k=1}^{N_e}\sum_{l=1}^{N_e}\left\{ \left[ \left( (\omega-\omega_0)+i\frac{\gamma_0}{2}\right)\mathbb{\mathbf{I}}+\frac{k_0^2}{\hbar \epsilon_0}\mathbf{K}(\omega)\right]^{-1}\right\}_{kl} \right \vert^2 \,,\\
&&D_{class}(\omega) \propto \left \vert \sum_{k=1}^{N_e}\sum_{l=1}^{N_e}\left\{ \left[(\omega-\omega_0)\mathbb{\mathbf{I}}+i\frac{\gamma_0}{2}\left(\mathbb{\mathbf{I}}+\frac{6\pi}{k_0}\mathbf{K}(\omega)\right)\right]^{-1}\right\}_{kl} \right \vert^2 \,,
\end{eqnarray}
\end{subequations}
where we have used $\gamma_0 =\frac{d^2\omega_0^3}{3\pi\epsilon_0 \hbar c^3}$. Therefore the eigenvalue of the classical model (Eq. \ref{eq:classic}) corresponds to the peak of the quantum model if we neglect the $\Delta \omega_c^{(ij)}$ contribution. $\Delta \omega_c^{(ij)}$ represents quantum corrections to the classical Drude Lorentz model.  Stated differently, the continuous model is equivalent to the classical model when the quantum corrections in the Lamb shift or dipole-dipole coupling are neglected. See also ref. \cite{Wylie-Sipe:1985b} for a deep discussion of quantum and classical descriptions of the Lamb shift. 

%\bibliography{/Users/gcolas/Desktop/ARTICLE/refFull,/Users/gcolas/Desktop/ADMINISTRATIF/CV/CVtex0/DONNEES/refGCF}

\end{document}